\documentclass[12pt]{article}
\usepackage{amsmath,amsfonts,amssymb,amsbsy}
\usepackage{natbib}
\usepackage[cp1251]{inputenc}
\usepackage[english]{babel}
\usepackage{graphicx}
\usepackage{color}
\long\def\symbolfootnote[#1]#2{\begingroup%
\def\thefootnote{\fnsymbol{footnote}}\footnote[#1]{#2}\endgroup}
\begin{document}

\centerline{\large\bf
Impact problem for the quasi-linear viscoelastic}
\centerline{\large\bf
standard solid model}
\medskip
\centerline{I.I.~Argatov, N.S.~Selyutina and G.S.~Mishuris\thanks{Corresponding author; e-mail:
{\tt ggm@aber.ac.uk}}}
\medskip
\centerline{\it Department of Mathematics, Aberystwyth University, Wales, UK}
\bigskip
\medskip

{\bf Abstract:}
The one-dimensional impact problem in the case of Fung's quasi-linear viscoelastic model is studied for the relaxation function of the standard solid model (or Zener model). At that, quasi-linear viscoelastic Maxwell and Kelvin--Voigt models are recovered as limit cases. The results of numerical simulations for some illustrative values of the dimensionless problem parameters are presented.
\medskip


\bigskip

\setcounter{equation}{0}
\section{Introduction}
\label{1dsSectionI}

Impact testing of soft biomedical materials and biological tissues represents a considerable practical interest \citep{Kalcioglu2011}.
In recent years there has also been a growing interest in impact testing at nanoscale \citep{Constantinides2008}.
The mechanical behavior of articular cartilage as a biological tissue demonstrates several complex features including viscoelasticity and nonlinearity \citep{HayesMockros1971}, and there have been developed a number of mathematical models for soft biological tissues and biomaterials, including poroelasticity \citep{MowKueiLaiArmstrong1980}, poroviscoelasticity \citep{SettonZhuMow1993}, and
rheological network modelling \citep{BischoffArrudaGrosh2004}.
Due to the short time span involved in the impact problem, disallowing macroscopic motion of the interstitial fluid relative to the solid matrix of the tissue, we ignore its multiphasic nature.
However, it should be noted that the validity of the assumption of viscoelasticity may be jeopardized if the indenter has a submicron diameter, because the macroscopic size of interest has been reduced to a very short distance.

Needless to say that the interpretation of experimental data on material characterization by high-rate impact tests crucially depends on the mathematical material model employed for the analysis.
Because biological materials exhibit time-dependent response to external mechanical stimuli, viscoelastic models seem to be appropriate candidates. The corresponding impact problem in spite of being simply formulated imposes considerable mathematical difficulties for analytical solution.
The analytical solution of one-dimensional impact problem is known for the Kelvin--Voigt  \citep{WinemanRajagopal2000,Popov2010} and Maxwell \citep{Stronge2000} models, and in the case of viscoelastic standard solid model it was studied by means of asymptotic methods \citep{ButcherSegalman2000,Ardatov2013}. Furthermore, some qualitative properties of the one-dimensional impact were established \citep{Ardatov2013} in the framework of general linear viscoelastic model represented by the Boltzmann's hereditary integral.

It is clear that linear impact models are not capable of describing sensitive features of impact phenomena for soft tissues \citep{Edelsten2010,Kalcioglu2011}. One of the most striking examples is the behavior of the coefficient of restitution as a function of the velocity of impactor observed in recent experiments on articular cartilage \citep{Varga2007,Edelsten2010}, whereas the linear viscoelastic impact models asserts that it should be independent of the impact velocity. That is why, it was hypothesized \citep{Edelsten2010,Ardatov2013} that utilizing non-linear viscoelastic models in the impact problem may shed light on the behavior of the output impact variables (e.g., coefficient of restitution) with variation of the input impact parameters (impactor mass and velocity).

In the present paper, we employ the quasi-linear viscoelastic (QLV) model, which was proposed by \citet{Fung1981} and is based on the hereditary integral with a certain relaxation function. In particular, we consider the following relaxation functions: 1) Standard solid model (Zener model), 2) Maxwell model, 3) Kelvin--Voigt model. At that, the QLV standard solid model is characterized by the following four parameters: Instantaneous elastic modulus, $E_0$, characteristic relaxation time, $\tau_R$, equilibrium-to-instantaneous modulus ratio, $\rho$, and (dimensionless) nonlinearity parameter, $B$.
Correspondingly, the QLV Maxwell and Kelvin--Voigt models are obtained from the QLV Zener model as limit cases as $\rho\to 0$ (after appropriate renormalization).
On the other hand, the QLV Fung model is an extension of the Boltzmann integral to account for elastic non-linearity via the parameter $B$ and therefore by definition
the QLV Zener model reduces to a standard solid model as $B\to 0$.

The paper is organized as follows. In Section\ref{1dsSection1}, we give the general background for the QLV standard solid model based one-dimensional impact problem. To facilitate numerical implementation, the impact problem is reduced a system of the first-order differential equations.
The two limit cases for the QLV Zener model are considered in Section~\ref{1dsSection3}.
The results of numerical simulations for some illustrative values of the dimensionless problem parameters are presented and discussed in Sections~\ref{1dsSectionB} and \ref{1dsSectionDC}, respectively.

\section{Impact problem formulation}
\label{1dsSection1}
\subsection{Equation of motion of the rigid impactor}
\label{1dsSub1section1}

\begin{figure}[h!]
\begin{center}
    \includegraphics[scale=0.32]{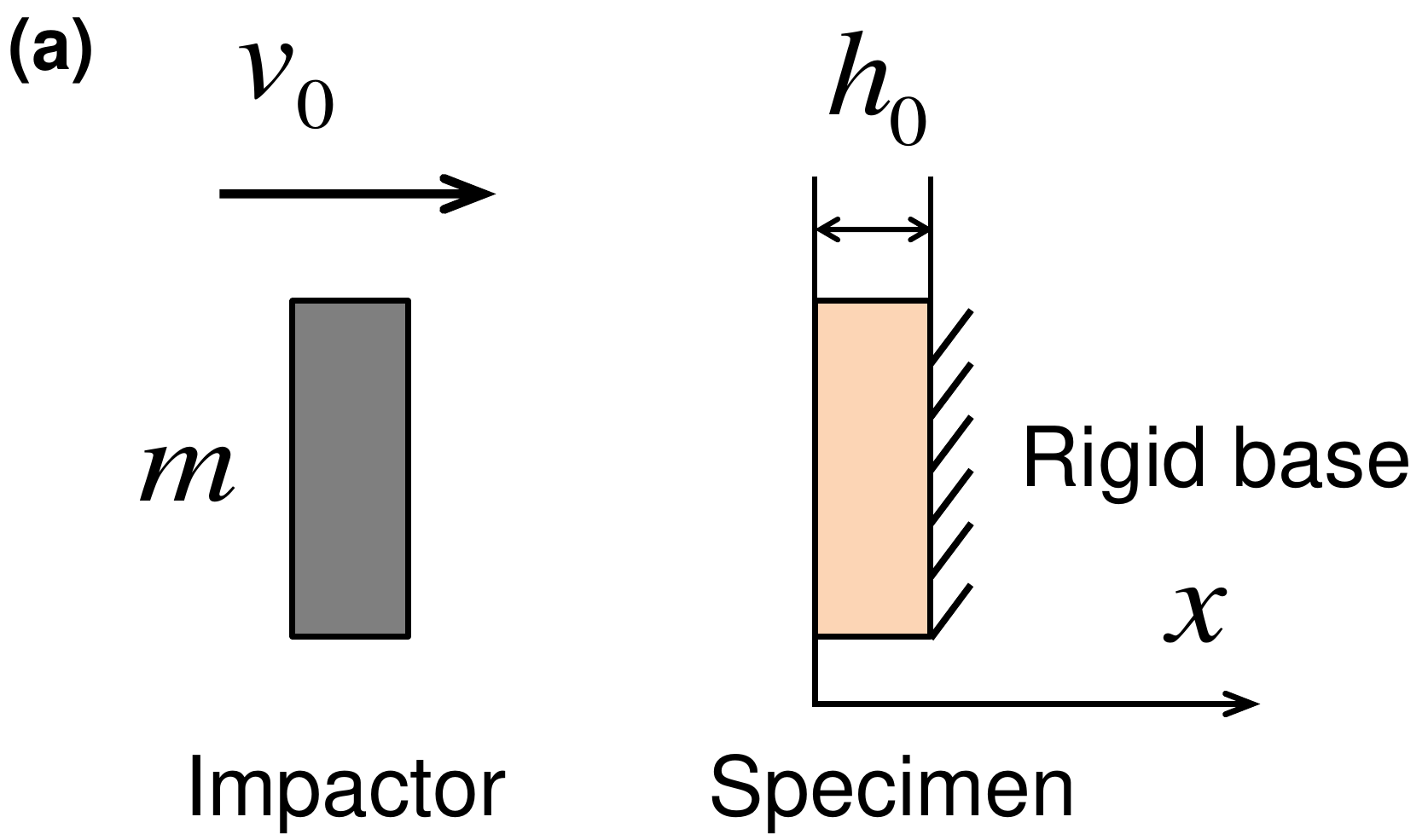}\hskip0.5cm
    \includegraphics[scale=0.32]{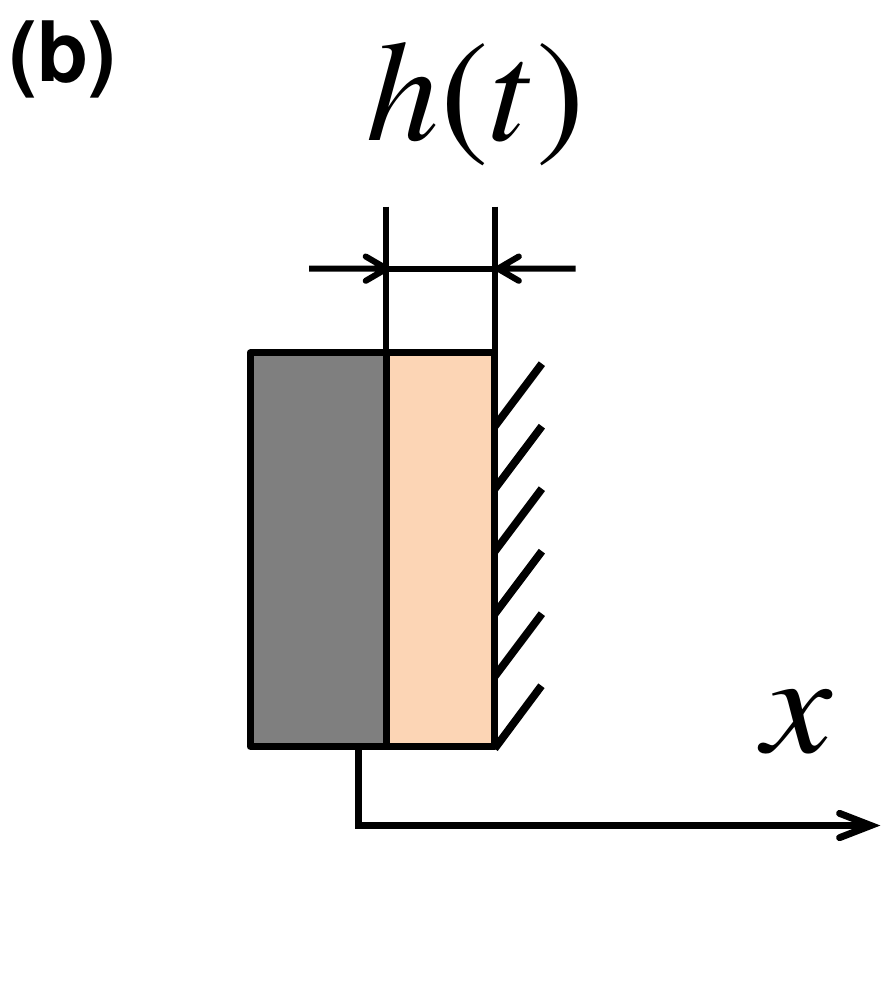}
\caption{Scheme of impact loading of the tissue specimen:
		(a) Initial configuration before the contact;
		(b) Current configuration.}
\end{center}
    \label{Schema.pdf}
\end{figure}

We consider the schematic representation for impact loading of a biological tissue specimen shown in Fig.~\ref{Schema.pdf}. On the basis of Newton's second law, the impactor motion is determined by the differential equation
\begin{equation}
m\ddot{x}=-F, \quad t\in(0,t_c),
\label{1vI(1.0)}
\end{equation}
with the initial conditions
\begin{equation}
x(0)=0, \quad \dot{x}(0)=v_0.
\label{1vI(1.1)}
\end{equation}

Here, $m$ is the mass of the impactor, $F$ is the reaction force of the tissue specimen (which is assumed to be positive), $v_0$ is the impactor speed at incidence, $t_c$ is the contact duration. In other words, $t_c$ denotes the instant when the specimen reaction force changes its sign and the impactor acceleration, $\ddot{x}$, vanishes, so that
\begin{equation}
F\bigr\vert_{t=t_c}=0.
\label{1vI(1.2)}
\end{equation}

An important characteristics of the impact problem (\ref{1vI(1.0)}), (\ref{1vI(1.1)}) is the coefficient of restitution, $e_*$, which is identified as the ratio of the absolute value of the impactor velocity at separation, $|\dot{x}(t_c)|$, to the impactor speed at the incidence, $\dot{x}(0)$, that is

\begin{equation}
e_*=\frac{|\dot{x}(t_c)|}{v_0}.
\label{1vI(1.3)}
\end{equation}

Moreover, some other impact characteristics also represent a considerable practical interest. Namely, the peak value, $x_m$, of the impactor displacement, which occurs at the time moment $t=t_m$ when $\dot{x}(t_m)=0$, and the peak value, $F_M$, of the contact force, which is achieved at the instant $t=t_M$, when $\dot{F}(t_M)=0$.

\subsection{Quasi-linear viscoelastic standard solid model}
\label{1dsSub2Section1}

We represent the specimen reaction force based on the quasi-linear viscoelastic (QLV) model introduced by \citet{Fung1981}. The QLV model was proposed for modeling biological tissues in one-dimensional stress state, which seems to be appropriate for blunt impact testing \citep{Edelsten2010,Varga2007}.
Let us recall that the one-dimensional stress, $\sigma(t)$, in the impacted specimen is determined as a ratio between the applied force $-F(t)$ and the undeformed cross-sectional area $A$. The corresponding extensional strain, $\varepsilon(t)$, is defined as
\begin{equation}
\varepsilon(t)=\frac{h(t)}{h_0}-1,
\label{1vI(2.0)}
\end{equation}
where $h(t)$ is the thickness of the tissue specimen at the current time, and $h_0=h(0)$.

According to the Fung QLV model, the dependence of the stress $\sigma(t)$ at time $t$ on the strain history is given by
\begin{equation}
\sigma(t)=\int\limits_{0^-}^t K(t-s)\frac{d\sigma^e}{d\varepsilon}\frac{d\varepsilon}{ds}(s)\,ds,
\label{1vI(2.1)}
\end{equation}
where it is assumed that $\varepsilon(t)=0$, $t\in(-\infty,0)$. The lower integration limit $0^-$ indicates that the integration in (\ref{1vI(2.1)}) starts at infinitesimally negative time so that to include the strain discontinuity at time zero.

Based on Fung's formula \citep{Fung1981,RajagopalSrivinasaWineman2007,MulianaRajagopal2012} for a nonlinear elastic stress function ${\sigma^e(\varepsilon(t))}$, the derivative $d\sigma^e/d\varepsilon$, appearing in Eq.~(\ref{1vI(2.1)}) is evaluated as
\begin{equation}
\frac{d\sigma^e}{d\varepsilon}=E_0\exp(B\varepsilon),
\label{1vI(2.3)}
\end{equation}
where $E_0$ and $B$ are material parameters. Note that the elastic modulus $E_0$ is related to the instantaneous elastic response of the tissue material.

Correspondingly, the stress-relaxation function, $K(t)$, which enters Eq.~(\ref{1vI(2.1)}) will be assumed to be normalized as
\begin{equation}
K(0)=1.
\label{1vI(2.2)}
\end{equation}

Due to geometric compatibility, the current thickness of the specimen, $h(t)$, and the initial thickness of the specimen, $h_0$, both entering Eq.~(\ref{1vI(2.0)}), are related to the impactor displacement, $x(t)$, through
\begin{equation}
h_0=h(t)+x(t).
\label{1vI(2.4)}
\end{equation}

Therefore, in view of (\ref{1vI(2.0)}), (\ref{1vI(2.3)}), and (\ref{1vI(2.4)}),  Eq.~(\ref{1vI(2.1)}) yields
\begin{equation}
F(x,t)=\frac{AE_0}{h_0}\int\limits_{0^-}^t K(t-s)\exp\Big(\frac{Bx(s)}{h_0}\Big)\frac{dx}{ds}(s)\,ds.
\label{1vI(2.5)}
\end{equation}

In the present paper, the normalized stress-relaxation function is taken according to the standard solid model, also known as the Zener model, as follows \citep{WinemanRajagopal2000}:
\begin{equation}
K(t)=\rho+(1-\rho)\exp\Big(-\frac{t}{\tau_R}\Big).
\label{1vI(2.K)}
\end{equation}
Here, $\rho$ represents the ratio between long-term and instantaneous responses, $\tau_R$ is a characteristic relaxation time.

\subsection{Reduction of the impact problem to a system of the first-order differential equations}
\label{1dsSection2}

In view of (\ref{1vI(2.5)}), Eq.~(\ref{1vI(1.0)}) can be recast in the form
\begin{equation}
m\ddot{x}(t)=-\frac{AE_0}{h_0}\int\limits_{0^-}^{t} K(t-s)\exp\Big(\frac{Bx(s)}{h_0}\Big)\dot{x}(s)\,ds.
\label{1vI(3.0)}
\end{equation}

Now, by introducing non-dimensional variables
\begin{equation}
\tau=\frac{t}{\tau_R}, \quad \xi=\frac{x}{v_0\tau_R},
\label{1vI(3.1)}
\end{equation}
the equation of motion (\ref{1vI(3.0)}) and the initial conditions (\ref{1vI(1.1)}) can be rewritten as follows:
\begin{equation}
\xi{''}+\alpha \mathcal{F}=0,
\label{1vI(3.2)}
\end{equation}
\begin{equation}
\xi(0)=0, \quad  \xi{'}(0)=1.
\label{1vI(3.3)}
\end{equation}
Here, the differentiation  with respect to the dimensionless time variable $\tau$ is denoted by stroke, $\mathcal{F}$ is the nondimensionalized reaction force given by
\begin{equation}
\mathcal{F}(\xi,\tau)=\int\limits_{0^-}^{\tau} \mathcal{K}(\tau-s)\exp(\beta\xi(s))\frac{d\xi}{ds}(s)\,ds,
\label{1vI(3.4)}
\end{equation}
\begin{equation}
\mathcal{K}(\tau)=\rho+(1-\rho)\exp(-\tau),
\label{1vI(3.5)}
\end{equation}
and we introduced the notation
\begin{equation}
\alpha=\frac{AE_0{\tau_R}^2}{mh_0}, \quad \beta=\frac{Bv_0\tau_R}{h_0}.
\label{1vI(3.6)}
\end{equation}

Furthermore, let us introduce an auxiliary notation
\begin{equation}
f(\xi,\tau)=\int\limits_{0^-}^{\tau} \exp(\beta\xi(s))\frac{d\xi}{ds}(s)\,ds,
\label{1vI(3.7)}
\end{equation}
\begin{equation}
\epsilon(\tau)=\frac{d\xi(\tau)}{d\tau}.
\label{1vI(3.8)}
\end{equation}

To proceed, first, we rewrite Eq.~(\ref{1vI(3.4)}) with (\ref{1vI(3.5)}), (\ref{1vI(3.7)}), and (\ref{1vI(3.8)}) taken into account as follows:
\begin{equation}
\mathcal{F}(\xi,\tau)=\rho f(\xi,\tau)+(1-\rho)\exp(-\tau)
\int\limits_{0^-}^{\tau}\exp(s)\exp(\beta\xi(s))\epsilon(s)\,ds.
\label{1vI(3.9)}
\end{equation}

Now, differentiating Eqs.~(\ref{1vI(3.7)}) and (\ref{1vI(3.9)}) with respect to $\tau$, we get
\begin{equation}
f{'}(\xi,\tau)=\epsilon(\tau)\exp(\beta\xi(\tau)),
\nonumber
\end{equation}
\begin{equation}
\mathcal{F}{'}(\xi,\tau)=\epsilon(\tau)\exp(\beta\xi(\tau))-\mathcal{F}(\xi,\tau)+\rho f(\xi,\tau).
\nonumber
\end{equation}

So, taking into account the above relations, we reduce the second order differential equation (\ref{1vI(3.2)}) to the following system of first-order differential equations:
\begin{equation}
    \begin {cases}
        \xi {'}=\epsilon,\\
        \epsilon{'}=-{\alpha}\mathcal{F},\\
        \mathcal{F}{'}=\epsilon\exp(\beta\xi)-\mathcal{F}+\rho f,\\
        f {'}=\epsilon\exp(\beta\xi).
    \end {cases}
\label{1vI(3.10)}
\end{equation}

The initial conditions (\ref{1vI(3.3)}) are supplemented with two new zero initial conditions for the new variables $\mathcal{F}$ and $f$, that is
\begin{equation}
\xi(0)=0, \quad \epsilon(0)=1, \quad \mathcal{F}(0)=0, \quad f(0)=0.
\label{1vI(3.11)}
\end{equation}

At that, in the non-dimensional variables (\ref{1vI(3.1)}), Eqs.~(\ref{1vI(1.2)}) and (\ref{1vI(1.3)}) for the contact duration, $t_c=\tau_R\tau_c$, and the coefficient restitution, respectively, take the form
\begin{equation}
\mathcal{F}\bigr\vert_{\tau=\tau_c}=0,
\label{1vI(3.12)}
\end{equation}
\begin{equation}
e_*=-\epsilon(\tau_c).
\label{1vI(3.13)}
\end{equation}

Note that the other impact characteristics introduced in Section~\ref{1dsSub1section1} can be evaluated as
\begin{equation}
t_m=\tau_R\tau_m,\quad x_m=v_0\tau_R\xi(\tau_m),
\label{1vI(3.13m)}
\end{equation}
\begin{equation}
t_M=\tau_R\tau_M,\quad
F_M=\frac{AE_0}{h_0}\mathcal{F}\bigr\vert_{\tau=\tau_M},
\label{1vI(3.13M)}
\end{equation}
where $\tau_m$ and $\tau_M$ are roots of the equations
\begin{equation}
\epsilon\vert_{\tau=\tau_m}=0,\quad
\mathcal{F}{'}\bigr\vert_{\tau=\tau_M}=0.
\label{1vI(3.13mM)}
\end{equation}

{
{\bf Remark 1.}
{\it
The obtained system (\ref{1vI(3.10)}) contains two dimensionless parameters $\alpha$ and $\beta$. In particular, the value of the parameter $\beta$ determines the extent of nonlinearity. So, if $\beta=0$, then the Cauchy problem (\ref{1vI(3.10)}), (\ref{1vI(3.11)}) reduces to the system
\begin{equation}
        \xi {'}=\epsilon,\quad
        \epsilon{'}=-{\alpha}\mathcal{F},\quad
        \mathcal{F}{'}=\epsilon-\mathcal{F}+\rho \xi,
\label{1vI(3.14)}
\end{equation}
\begin{equation}
\xi(0)=0, \quad \epsilon(0)=1, \quad \mathcal{F}(0)=0,
\nonumber
\end{equation}
which, in turn, results in the following differential equation of the third order:
\begin{equation}
\xi{'''}+\xi{''}+\alpha\xi{'}+\alpha\rho\xi=0,
\label{1vI(3.15)}
\end{equation}
subjected to the initial conditions
\begin{equation}
\xi(0)=0, \quad \xi{'}(0)=1, \quad \xi{''}(0)=0.
\label{1vI(3.15i)}
\end{equation}
The obtained Cauchy problem describes the linear impact problem for the Zener model studied in detail by \citet{ButcherSegalman2000,Ardatov2013}.
}}

\section{Limit cases for the QLV standard solid model}
\label{1dsSection3}

\subsection{QLV Maxwell model}
\label{1dsSub1section3}

By passing to the limit as $\rho\to 0$ in formula (\ref{1vI(2.K)}) for the stress-relaxation function of the Zener model, we obtain the relaxation function for viscoelastic Maxwell model,
$K_{\rm M}(t)$, which in the non-dimensional variables (\ref{1vI(3.1)}) takes the form
\begin{equation}
\mathcal{K}_{\rm M}(\tau)=\exp(-\tau).
\label{1vI(4.1)}
\end{equation}

Correspondingly, formula (\ref{1vI(2.5)}) yields the following expression for the specimen reaction force:
\begin{equation}
F_{\rm M}(x,t)=k_0\mathcal{F}_{\rm M}(\xi,\tau),
\label{1vI(4.FM)}
\end{equation}
where
\begin{equation}
k_0=\frac{AE_0}{h_0},
\label{1vI(4.k0)}
\end{equation}
\begin{equation}
\mathcal{F}_{\rm M}(\xi,\tau)=\exp(-\tau)\int\limits_{0^-}^{\tau}\exp(s) \exp(\beta\xi(s))\frac{d\xi}{ds}(s)\,ds.
\label{1vI(4.4)}
\end{equation}

Similarly to Section~\ref{1dsSection2}, it can be shown that the impact problem
\begin{equation}
\xi{''}+\alpha \mathcal{F}_{\rm M}=0,\quad
\xi(0)=0, \quad  \xi{'}(0)=1
\label{1vI(4.3)}
\end{equation}
is equivalent to the system
\begin{equation}
    \begin {cases}
        \xi {'}=\epsilon,\\
        \epsilon{'}=-{\alpha}\mathcal{F}_{\rm M},\\
        \mathcal{F}_{\rm M}{'}=\epsilon\exp(\beta\xi)-\mathcal{F}_{\rm M}
    \end {cases}
\label{1vI(4.10M)}
\end{equation}
with the initial conditions
\begin{equation}
\xi(0)=0, \quad \epsilon(0)=1, \quad \mathcal{F}_{\rm M}(0)=0.
\label{1vI(4.11I)}
\end{equation}
Note that the system (\ref{1vI(4.10M)}) can be obtained from (\ref{1vI(3.10)}) by passing to the limit  $\rho=0$.

In view of (\ref{1vI(1.2)}) and (\ref{1vI(1.3)}), the non-dimensional contact duration, $\tau_c$, and the coefficient restitution, $e_*$, are determined by the equations
\begin{equation}
\mathcal{F}_{\rm M}\bigr\vert_{\tau=\tau_c}=0,\quad
e_*=-\epsilon(\tau_c).
\label{1vI(4.9)}
\end{equation}

{
{\bf Remark 2.}
{\it
In the limit case $\beta=0$, the impact problem (\ref{1vI(4.10M)}), (\ref{1vI(4.11I)}) reduces to the Cauchy problem
\begin{equation}
\xi {'}=\epsilon,\quad
\epsilon{'}=-{\alpha}\mathcal{F}_{\rm M},\quad
\mathcal{F}{'}_{\rm M}=\epsilon-\mathcal{F}_{\rm M},
\label{1vI(4.6)}
\end{equation}
\begin{equation}
\xi(0)=0, \quad v(0)=1, \quad \mathcal{F}_{\rm M}(0)=0.
\label{1vI(4.7)}
\end{equation}
In this case, the coefficient restitution is given by the following formula \citep{ButcherSegalman2000,Stronge2000}:
\begin{equation}
e_*=\exp\Big(\frac{-\pi}{\sqrt{4\alpha-1}}\Big).
\label{1vI(4.9M)}
\end{equation}
Note also that the quantity $1/(2\sqrt{\alpha})$ has a physical meaning of the loss factor in the Maxwell model, which is schematically modeled as the combination of a spring (with the stiffness $k_0$ given by
(\ref{1vI(4.k0)})) and a dashpot (with the damping constant $b=\tau_R k_0$) in serial connection.
}}

\subsection{QLV Kelvin--Voigt model}
\label{1dsSub1section4}

First of all, let us rewrite formula (\ref{1vI(2.K)}) in the form
\begin{equation}
E_0 K(t)=E_\infty\Bigl\{1+\frac{1-\rho}{\rho}
\exp\Big(-\frac{t}{\tau_R}\Big)\Bigr\},
\label{1vI(2.Kinfty)}
\end{equation}
where the elastic modulus $E_\infty$ is related to the equilibrium elastic response of the tissue material.

In the framework of the standard solid model, the following relation holds between $E_\infty$ and $E_0$:
$$
E_\infty=\rho E_0.
$$

Let us also introduce the so-called retardation time by
$$
\tau_R^\prime=\frac{\tau_R}{\rho}.
$$

Therefore, that Eqs.~(\ref{1vI(2.5)}) and (\ref{1vI(2.K)}) can be recast as follows:
\begin{equation}
F(x,t)=\frac{AE_\infty}{h_0}\int\limits_{0^-}^t K^\prime(t-s)\exp\Big(\frac{Bx(s)}{h_0}\Big)\frac{dx}{ds}(s)\,ds,
\nonumber
\end{equation}
\begin{equation}
K^\prime(t)=1+\frac{1-\rho}{\rho}
\exp\Big(-\frac{t}{\rho\tau_R^\prime}\Big).
\nonumber
\end{equation}

Now, passing to the limit as $\rho\to 0$, the above formulas lead to the quasi-linear Kelvin--Voigt model:
\begin{equation}
F_{\rm KV}(x,t)=\frac{AE_\infty}{h_0}\int\limits_{0^-}^t
K_{\rm KV}(t-s)\exp\Big(\frac{Bx(s)}{h_0}\Big)\frac{dx}{ds}(s)\,ds,
\label{1vI(2.5KV)}
\end{equation}
\begin{equation}
K_{\rm KV}(t)=1+\delta\Big(\frac{t}{\tau_R^\prime}\Big).
\label{1vI(5.0)}
\end{equation}
Here, $\delta(x)$ is the Dirac delta function.

Thus, by substituting (\ref{1vI(5.0)}) into Eq.~(\ref{1vI(2.5KV)}), we arrive at the relation
\begin{equation}
F_{\rm KV}(x,t)=\frac{AE_\infty}{h_0}\biggl\{
\frac{h_0}{B}\Bigl[\exp\Big(\frac{Bx(t)}{h_0}\Big)-1\Bigr]
+\tau_R^\prime\exp\Big(\frac{Bx(t)}{h_0}\Big)\dot{x}(t)\biggr\}.
\label{1vI(5.1)}
\end{equation}

Further, introducing the non-dimensional variables
\begin{equation}
z=\exp\Big(\frac{Bx(t)}{h_0}\Big)-1,\quad
\tau=\frac{t}{\tau_R^\prime},
\label{1vI(5.2)}
\end{equation}
we reduce the impact problem (\ref{1vI(1.0)}), (\ref{1vI(1.1)}), (\ref{1vI(5.1)}) to the following Cauchy problem:
\begin{equation}
z{''}-\frac{z{'}^2}{z+1}=-\alpha^\prime(z+z{'})(z+1),
\label{1vI(5.4)}
\end{equation}
\begin{equation}
z(0)=0, \quad z{'}(0)=\beta^\prime.
\label{1vI(5.5)}
\end{equation}
Here we introduced the notation
\begin{equation}
\alpha^\prime=\frac{AE_\infty{\tau_R}^{\prime 2}}{mh_0},\quad
\beta^\prime=\frac{Bv_0\tau_R^\prime}{h_0}.
\label{1vI(3.6prime)}
\end{equation}

According to (\ref{1vI(1.2)}) and(\ref{1vI(1.3)}), the contact contact duration,
$t_c=\tau_R^\prime\tau_c$, and the coefficient restitution are identified as
\begin{equation}
z+z{'}\bigr\vert_{\tau=\tau_c}=0,
\label{1vI(5.6)}
\end{equation}
\begin{equation}
e_*=\frac{1}{\beta^\prime}\frac{|z{'}|}{z+1}\Biggr\vert_{\tau=\tau_c}.
\label{1vI(5.7)}
\end{equation}

Finally, the peak value of the impactor displacement, $x_m$, which occurs at the time moment $t_m=\tau_R^\prime\tau_m$, and the peak value of the contact force, $F_M$, which occurs at the time moment $t_M=\tau_R^\prime\tau_M$, can be evaluated as follows:
\begin{equation}
z^\prime\bigr\vert_{\tau=\tau_m}=0,\quad x_m=\frac{h_0}{B}\ln(z(\tau_m)+1),
\label{1vI(5.13m)}
\end{equation}
\begin{equation}
z^\prime+z^{\prime\prime}\bigr\vert_{\tau=\tau_M}=0,\quad
F_M=\frac{AE_\infty}{B}(z(\tau_M)+z^\prime(\tau_M)).
\label{1vI(5.13M)}
\end{equation}

{
{\bf Remark 3.}
{\it
Note that the change of the variable $(\ref{1vI(5.2)})_1$ does not allow us to pass to the limit $\beta^\prime=0$ in the impact problem (\ref{1vI(5.6)}), (\ref{1vI(5.7)}). However, in the limit case $\beta^\prime=0$ (that is when $B=0$), the impact problem under consideration reduces to the following Cauchy problem (see formula (\ref{1vI(5.1)})), which corresponds to the linear viscoelastic Kelvin--Voigt model:
\begin{equation}
m\ddot{x}=-\frac{AE_\infty}{h_0}\bigl\{x(t)+\tau_R^\prime\dot{x}(t)\bigr\}, \quad t\in(0,t_c),
\nonumber
\end{equation}
\begin{equation}
x(0)=0, \quad \dot{x}(0)=v_0.
\nonumber
\end{equation}
In this case, the coefficient restitution is given by the following formula \citep{ButcherSegalman2000,Popov2010,WinemanRajagopal2000}:
\begin{equation}
e_*=\exp\Bigg(-\frac{2\sqrt{\alpha^\prime}}{\sqrt{4-\alpha^\prime}}{\,\rm atan\,}
\frac{\sqrt{4-\alpha^\prime}}{\sqrt{\alpha^\prime}}\Bigg),
\label{1vI(5.12)}
\end{equation}
where $\sqrt{\alpha^\prime}/2$ is the loss factor in the Kelvin--Voigt model, which is schematically modeled as a parallel combination of a linear spring with the stiffness
$k_\infty=AE_\infty/h_0$ and a dashpot with the damping coefficient $b=k_\infty\tau_R^\prime$.
}}

\section{Behavior of the main impact parameters}
\label{1dsSectionB}

First of all, to illustrate numerical solutions of the impact problems for the QLV Maxwell and Kelvin-Voigt models, we consider the following ranges for the model dimensionless parameters:
$\beta$  and $\beta^\prime$ from 0 to 15, $\alpha$ from 0.25 to 100, and $\alpha^\prime$ from 0 to 4.
Note that these parameters, which are introduced in (\ref{1vI(3.6)}) and (\ref{1vI(3.6prime)}), can be viewed as non-dimensionalized reciprocal mass of the impactor and its initial velocity.

In all figures below, the following convention is used: (0) dot-dash black line represents the limiting case for the corresponding second parameter, while the  solid lines (1) red, (2) dark blue, (3) purple, and (4) green are plotted in the ascending order of this parameter.

\begin{figure}[h!]
    \centering
    \includegraphics [scale=0.30]{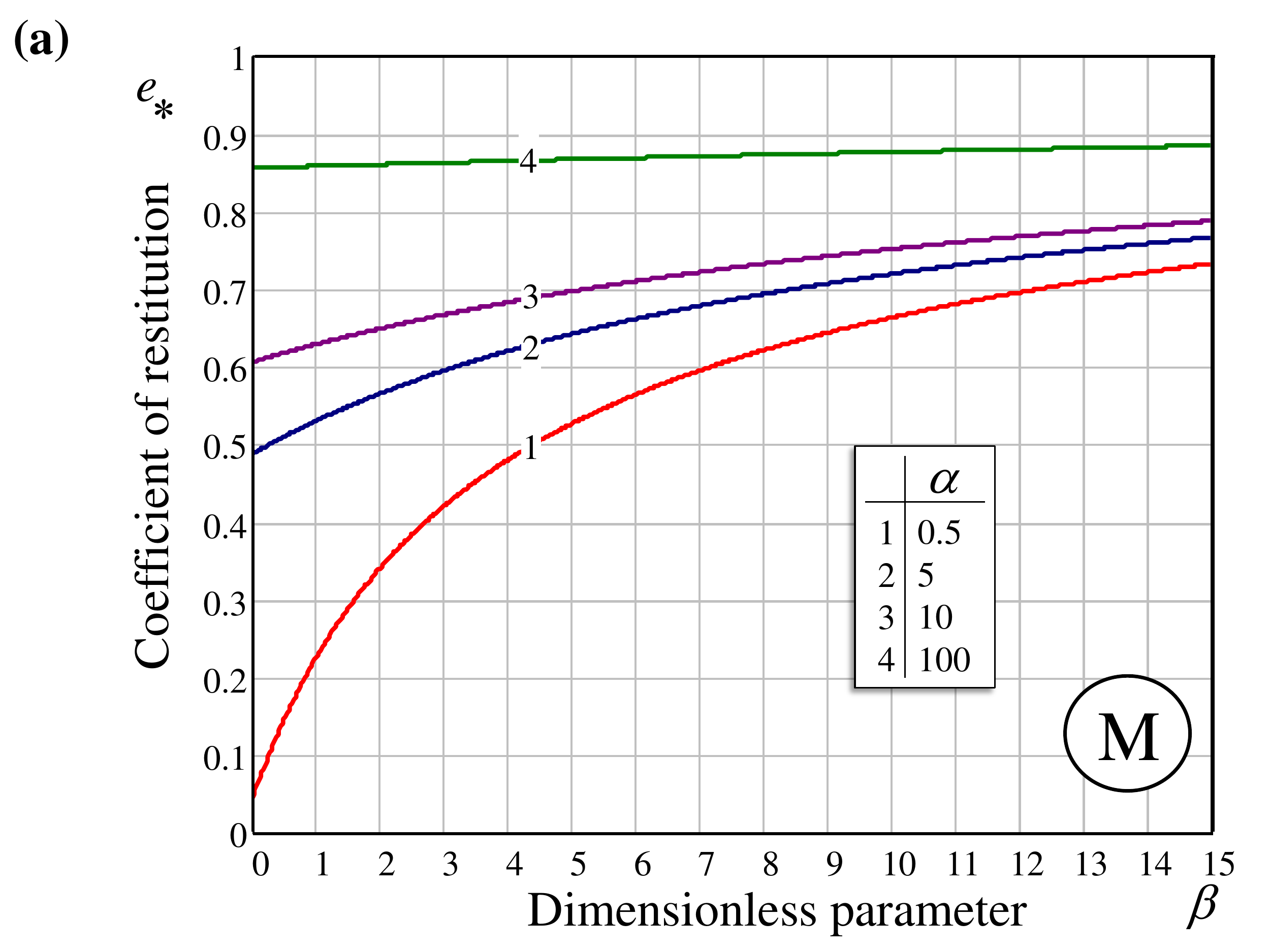}
    \includegraphics [scale=0.30]{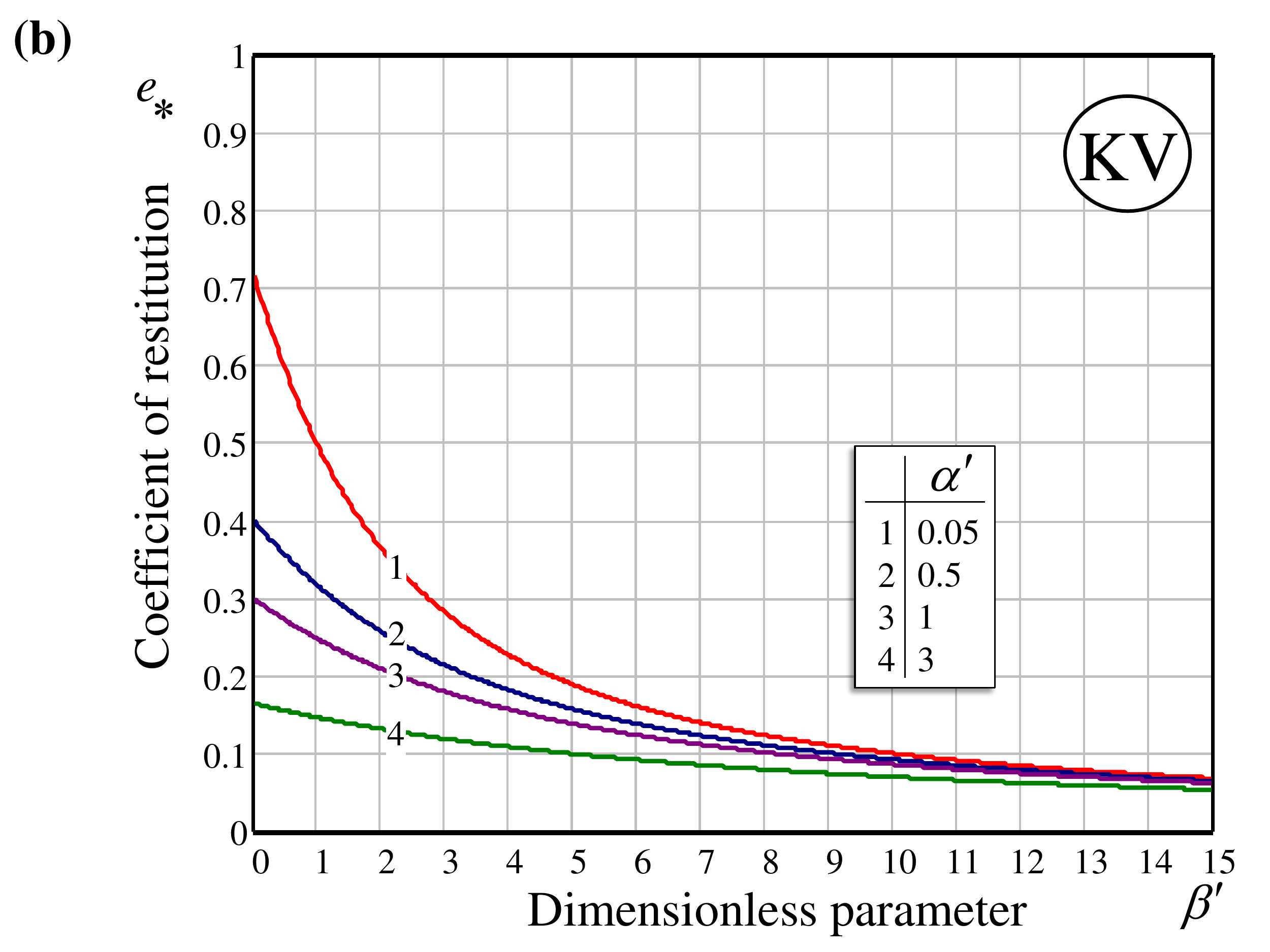}
        \caption{Coefficient of restitution as a function of the relative incident velocity:
         (a) QLV Maxwell model, $\beta=Bv_0\tau_R/h_0$; (b) QLV Kelvin--Voigt model, $\beta^\prime=Bv_0\tau_R^\prime/h_0$.}
    \label{eQLVbeta.pdf}
\end{figure}

Fig.~\ref{eQLVbeta.pdf} shows the behavior of the coefficient of restitution as a function of the relative incident velocity for (a) QLV Maxwell and (b) Kelvin-Voigt models, respectively. It is to note that with the increase of the relative incident velocity, the coefficient of restitution increases for the QLV Maxwell model and decreases for the QLV Kelvin-Voigt model. Recall that for the linear Maxwell and Kelvin-Voigt models, the coefficient of restitution is independent of the incident velocity.

\begin{figure}[h!]
    \centering
    \includegraphics [scale=0.30]{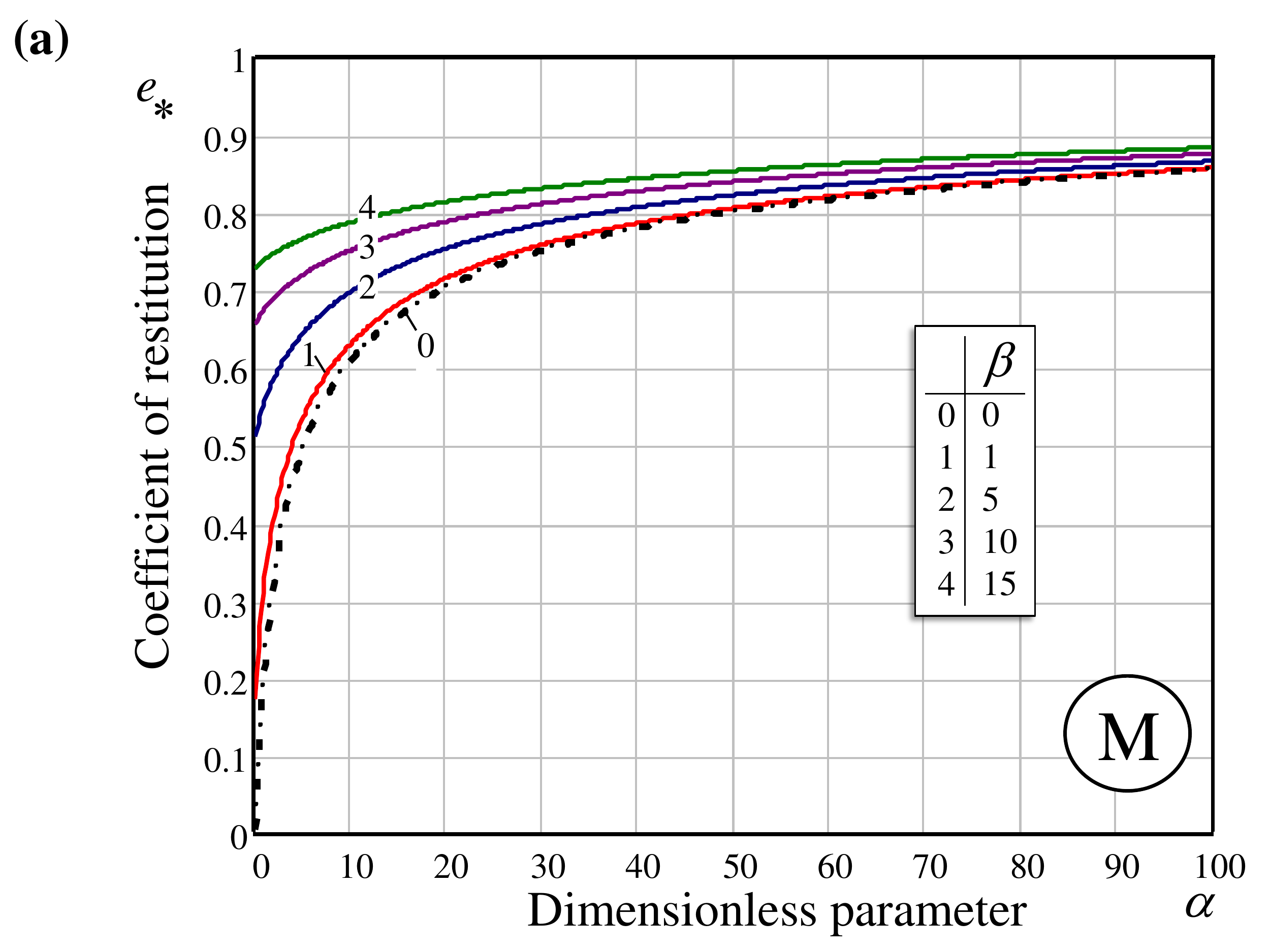}
    \includegraphics [scale=0.30]{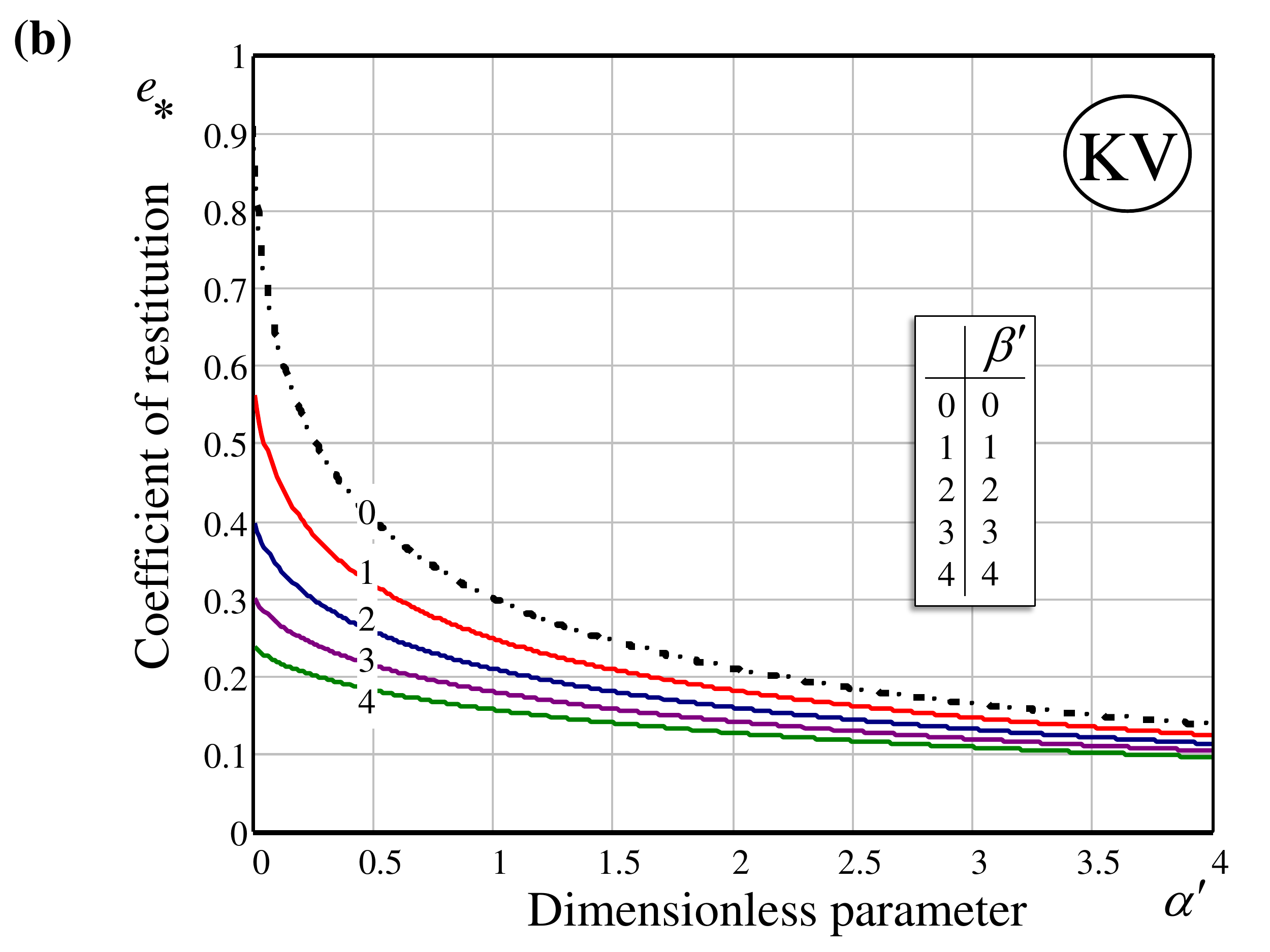}
        \caption{Coefficient of restitution as a function of the reciprocal relative mass of the impactor:
         (a) QLV Maxwell model, $\alpha=AE_0\tau_R^2/(mh_0)$; (b) QLV Kelvin--Voigt model,
$\alpha^\prime=AE_\infty\tau_R^{\prime 2}/(mh_0)$.}
    \label{eQLValpha.pdf}
\end{figure}

The behavior of the restitution coefficient as a function of the reciprocal relative impactor mass for (a) QLV Maxwell and (b) QLV Kelvin-Voigt models is presented in Fig.~\ref{eQLValpha.pdf}. It is readily seen that with the increase of the relative mass of the impactor, the coefficient of restitution increases for the QLV Maxwell model and decreases for the QLV Kelvin-Voigt model. Observe that this general tendency for the coefficient of restitution is in complete agreement with the behavior of the restitution coefficient in the limiting cases, $\beta=0$ and $\beta^\prime=0$, as it follows from formulas (\ref{1vI(4.9M)}) and (\ref{1vI(5.12)}).

\begin{figure}[h!]
    \centering
    \includegraphics [scale=0.30]{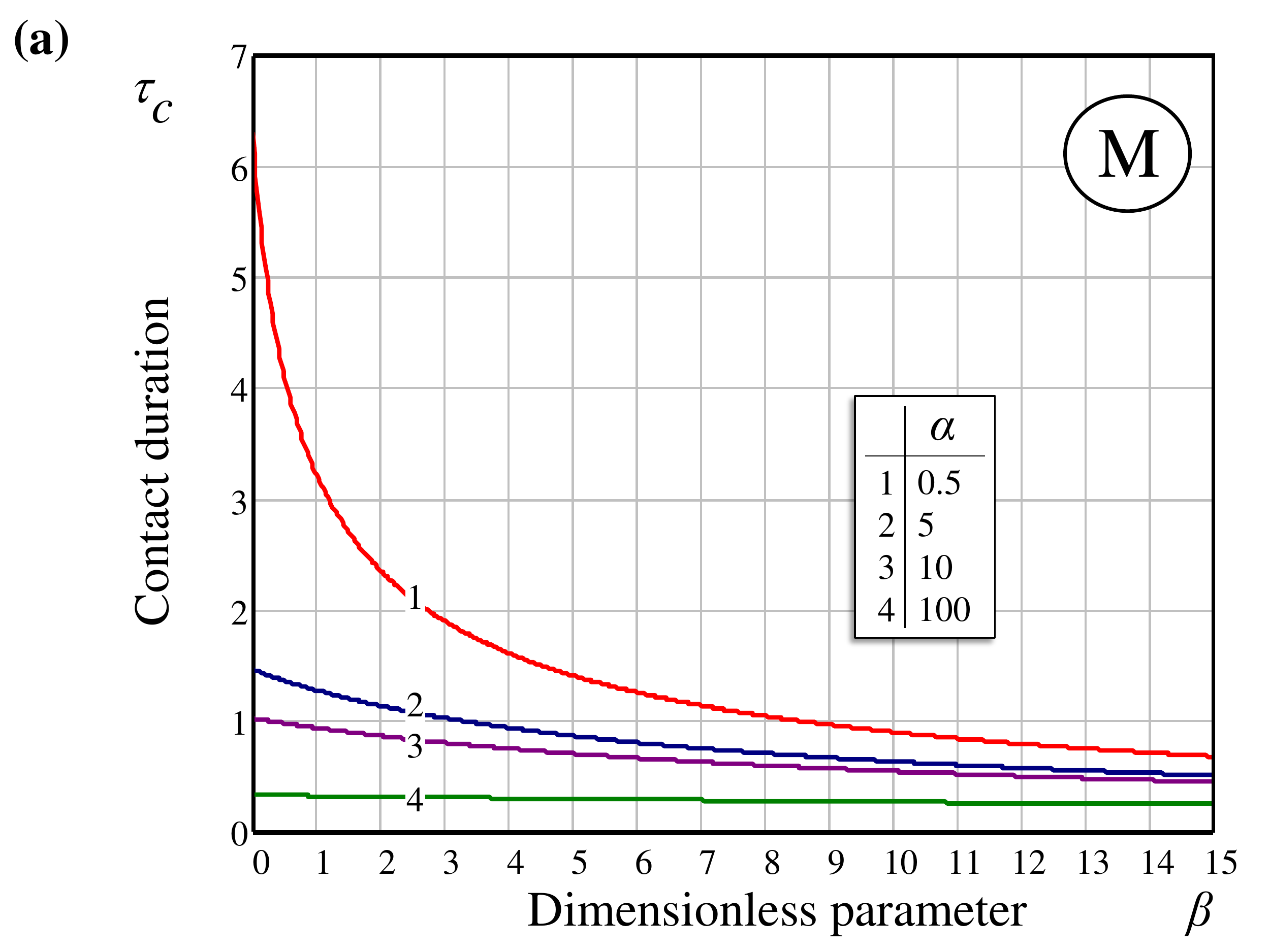}
    \includegraphics [scale=0.30]{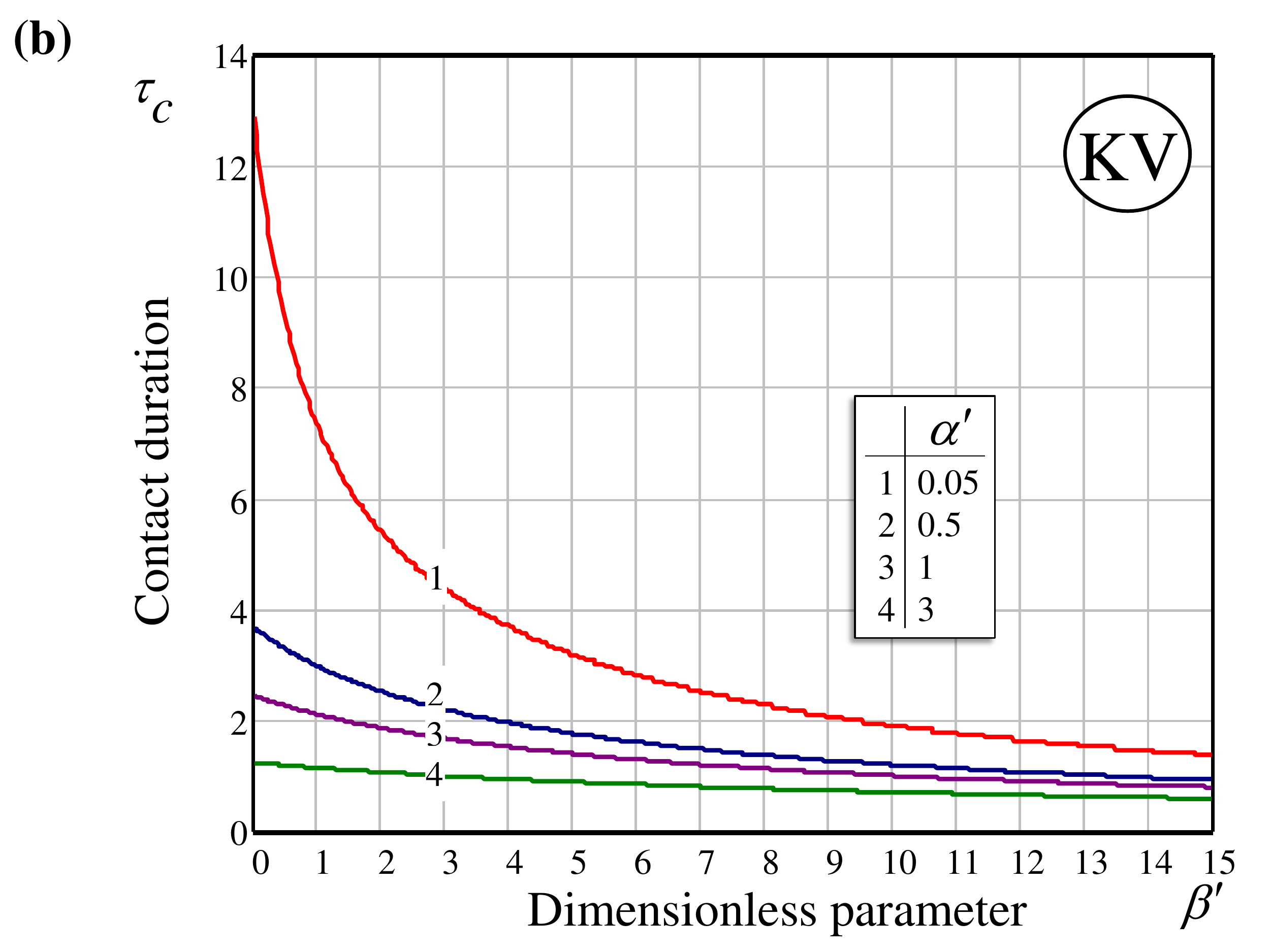}
        \caption{Relative contact duration as a function of the relative incident velocity:
         (a) QLV Maxwell model, $\tau_c=t_c/\tau_R$ and $\beta=Bv_0\tau_R/h_0$; (b) QLV Kelvin--Voigt model, $\tau_c=t_c/\tau_R^\prime$ and $\beta^\prime=Bv_0\tau_R^\prime/h_0$.}
    \label{tau_cQLVbeta.pdf}
\end{figure}

\begin{figure}[h!]
    \centering
    \includegraphics [scale=0.30]{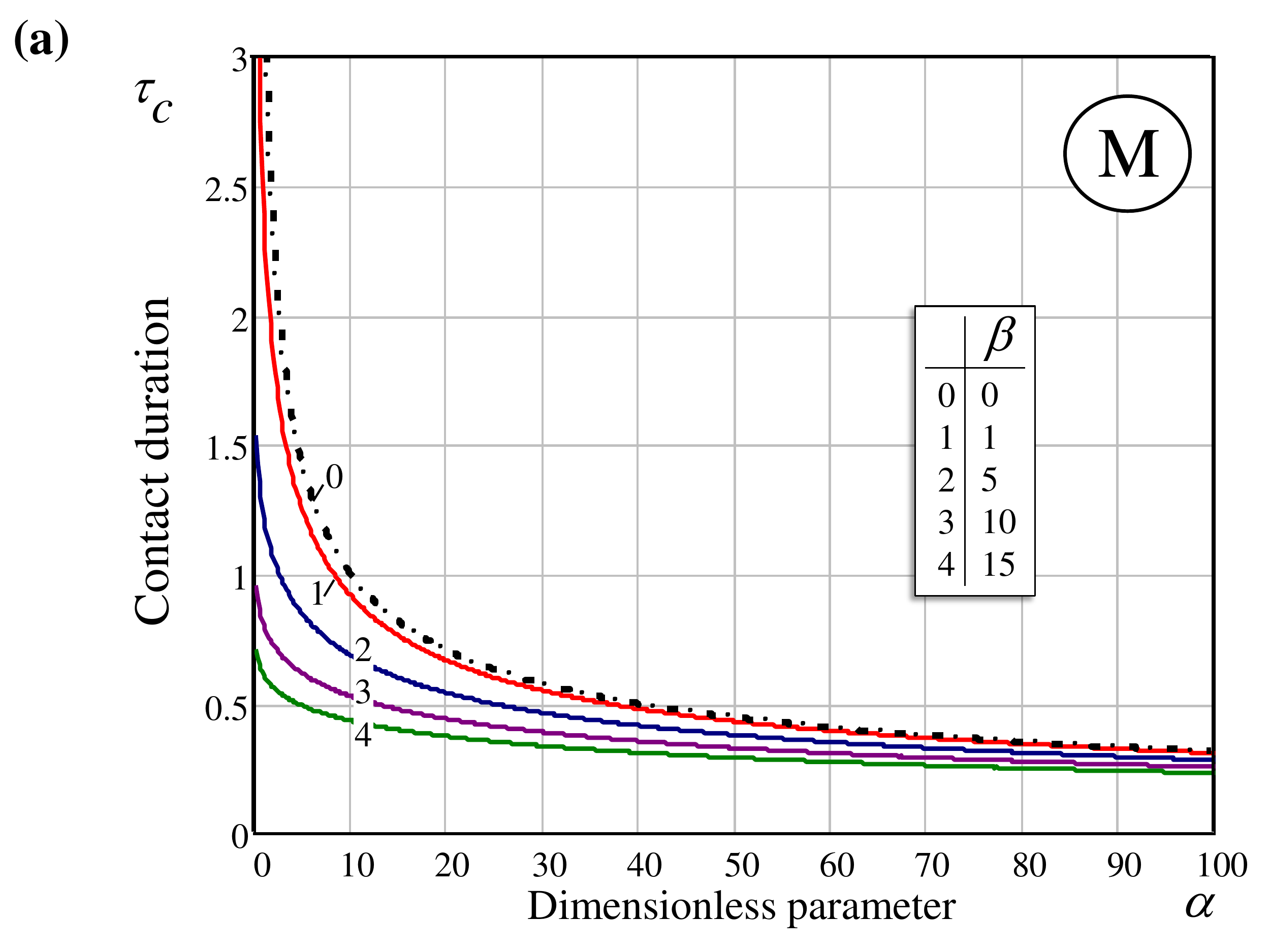}
    \includegraphics [scale=0.30]{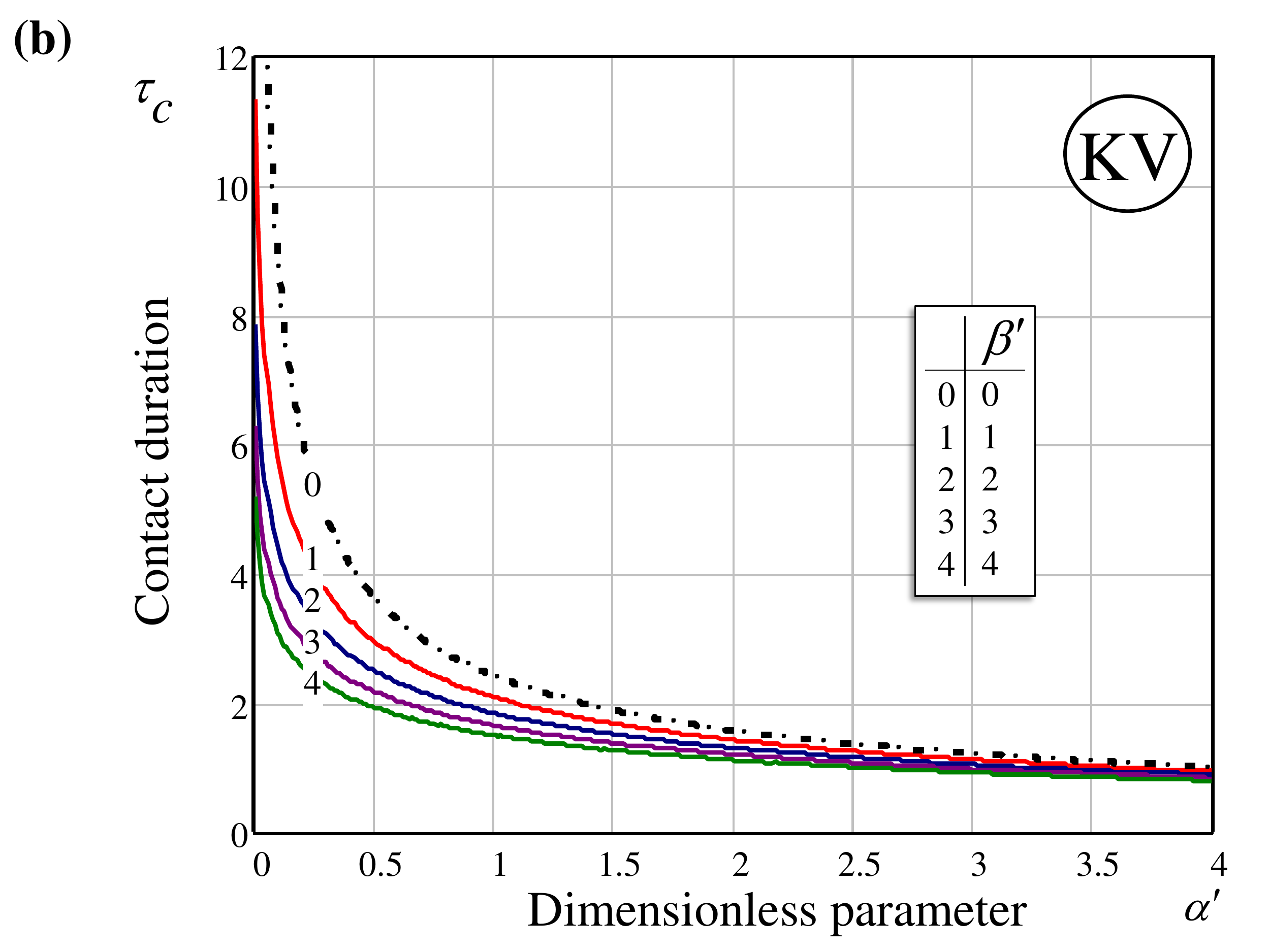}
        \caption{Relative contact duration as a function of the reciprocal relative mass of the impactor:
         (a) QLV Maxwell model, $\tau_c=t_c/\tau_R$ and $\alpha=AE_0\tau_R^2/(mh_0)$ (b) QLV Kelvin--Voigt model, $\tau_c=t_c/\tau_R^\prime$ and $\alpha^\prime=AE_\infty\tau_R^{\prime 2}/(mh_0)$.}
    \label{taucQLVMKV=04.pdf}
\end{figure}

In view of the opposite behavior of the the restitution coefficient in the QLV Maxwell and Kelvin-Voigt models, it makes sense to consider the behavior of the duration of the impact process when the contact between the impact and the tissue specimen takes place.
Fig.~\ref{tau_cQLVbeta.pdf} shows the behavior of the relative contact duration as function of the relative incident velocity for (a) QLV Maxwell and (b) QLV Kelvin-Voigt models. The dependence of the relative contact duration on the reciprocal relative impactor mass is plotted in Fig.~\ref{taucQLVMKV=04.pdf}. It is of interest to observe that the relative contact duration for the both QLV models decreases with the increase of either relative incident velocity or reciprocal relative impactor mass.

\begin{figure}[h!]
    \centering
    \includegraphics [scale=0.30]{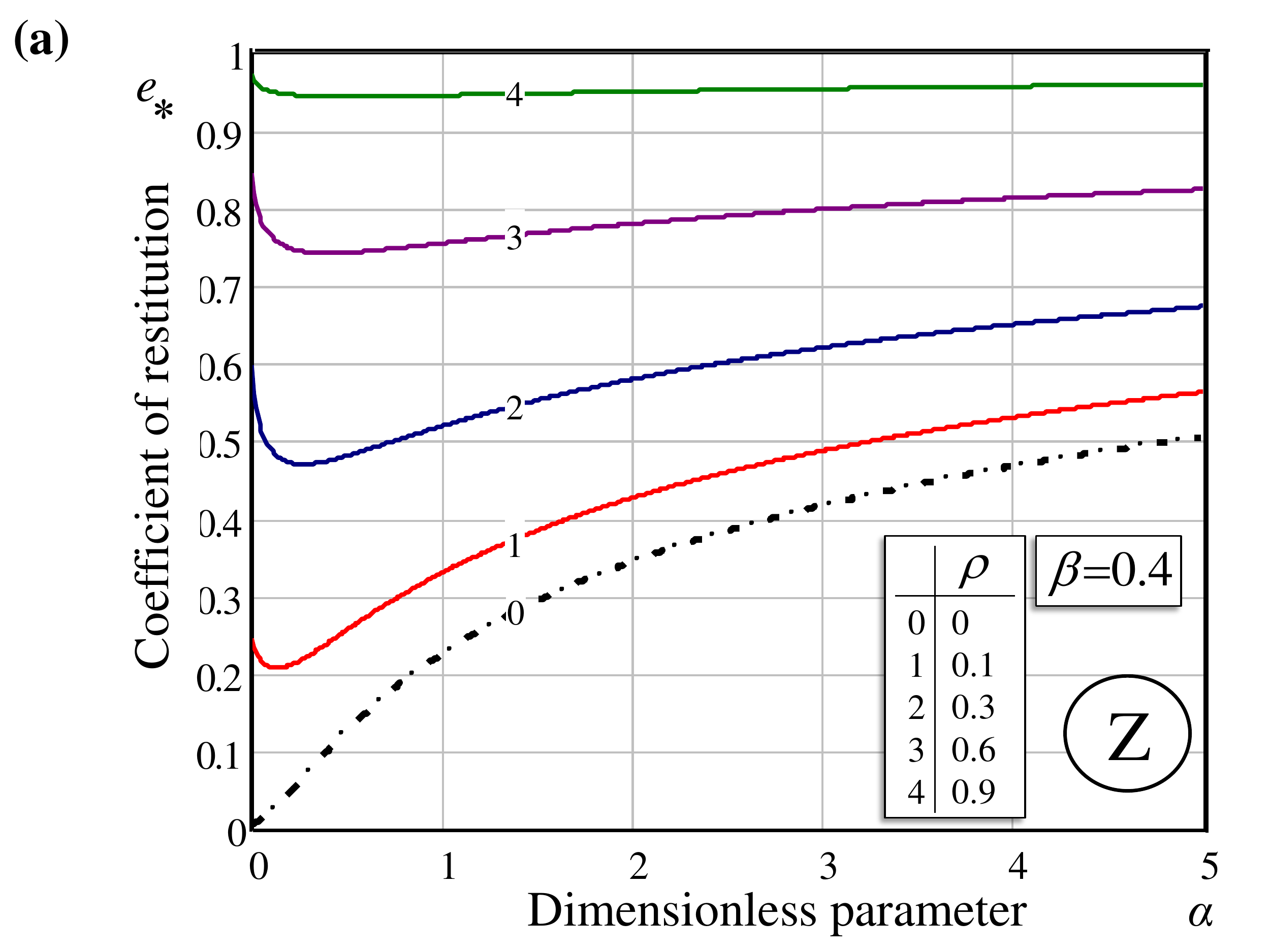}
    \includegraphics [scale=0.30]{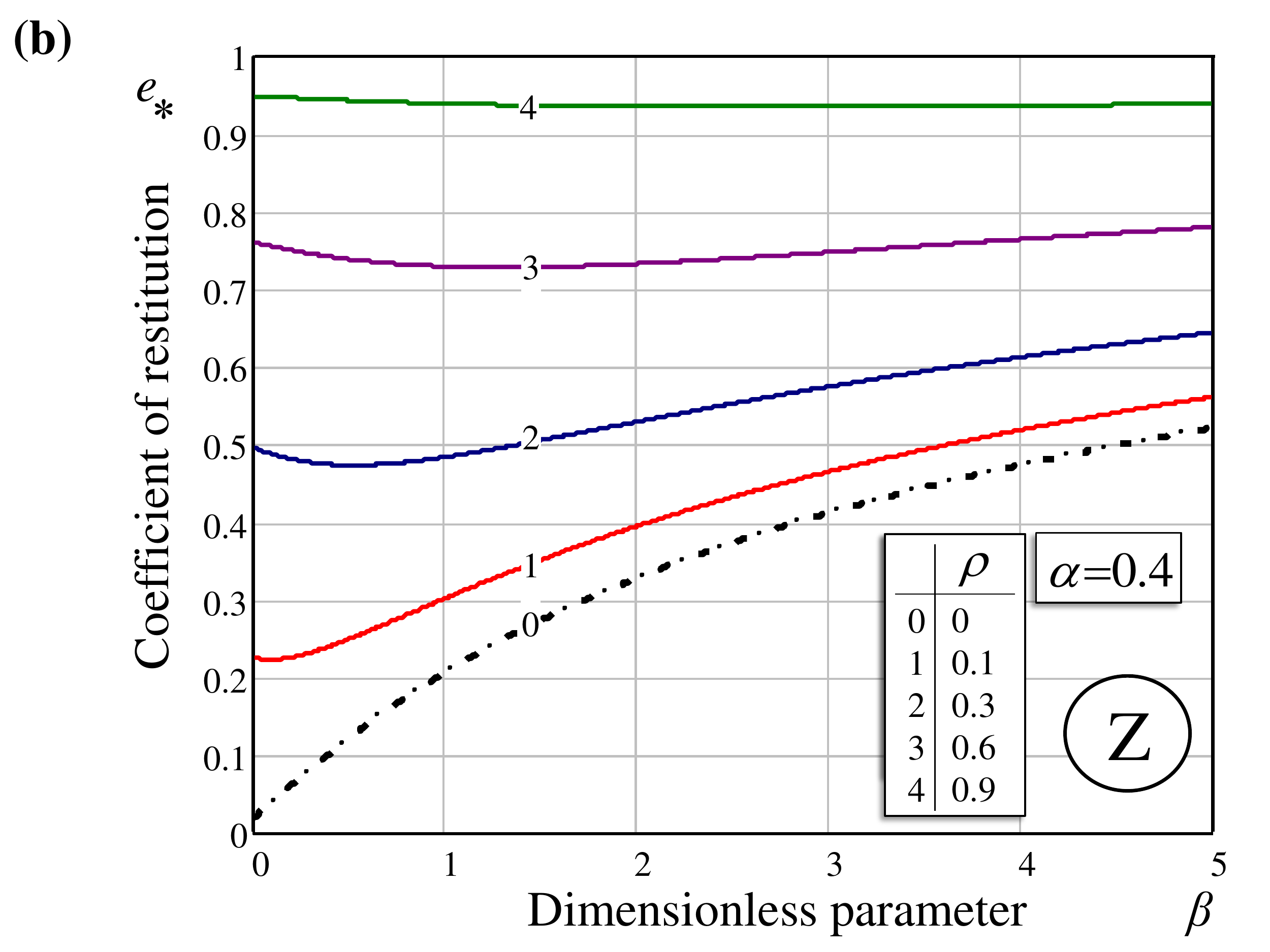}
        \caption{Coefficient of restitution in the QLV standard solid model as a function of:
(a) the reciprocal relative mass of the impactor $\alpha$ with the fixed relative incident velocity $\beta=0{.}4$ and
(b) the relative incident velocity $\beta$ with the fixed reciprocal relative mass of the impactor $\alpha=0{.}4$ for different values of the elastic moduli ratio $\rho$.}
    \label{eQLVSSM=04}
\end{figure}

\begin{figure}[h!]
    \centering
    \includegraphics [scale=0.30]{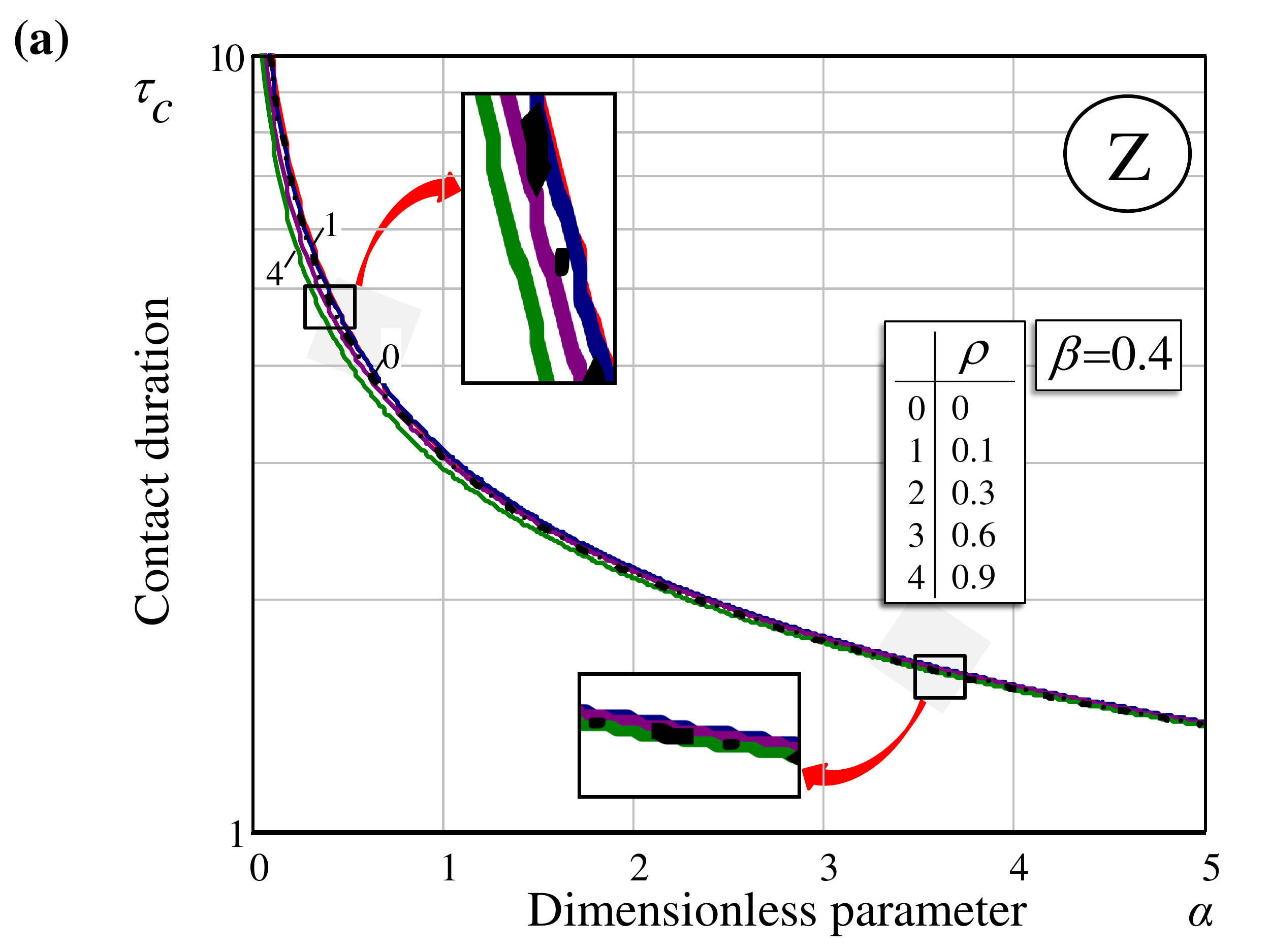}
    \includegraphics [scale=0.30]{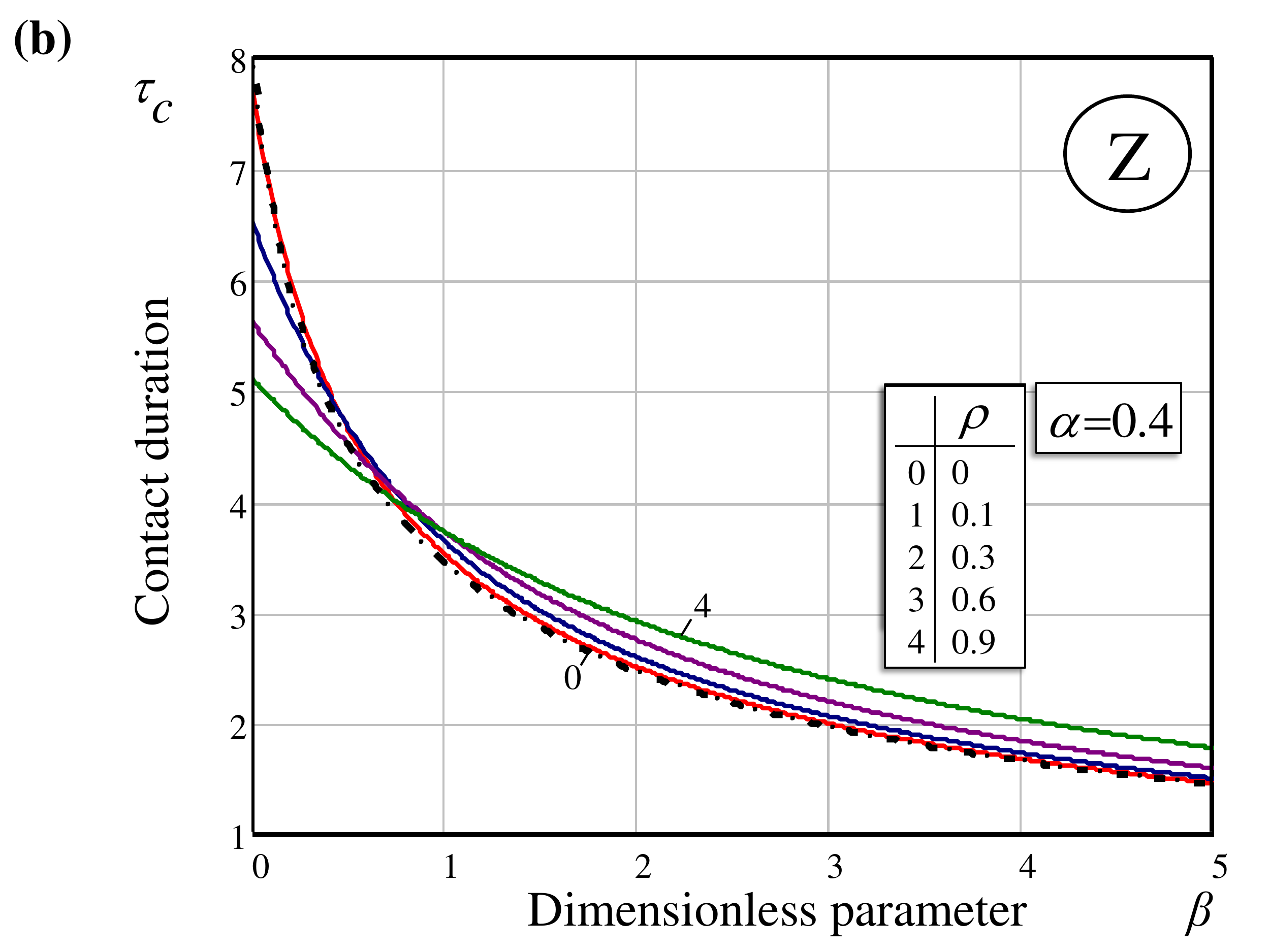}
        \caption{Relative contact duration $\tau_c=t_c/\tau_R$ in the QLV standard solid model as a function of:
(a) the reciprocal relative mass of the impactor $\alpha$ with the fixed relative incident velocity $\beta=0{.}4$ and
(b) the relative incident velocity $\beta$ with the fixed reciprocal relative mass of the impactor $\alpha=0{.}4$ for different values of the elastic moduli ratio $\rho$.}
    \label{taucQLVSSM=04}
\end{figure}

\begin{figure}[h!]
    \centering
    \includegraphics [scale=0.30]{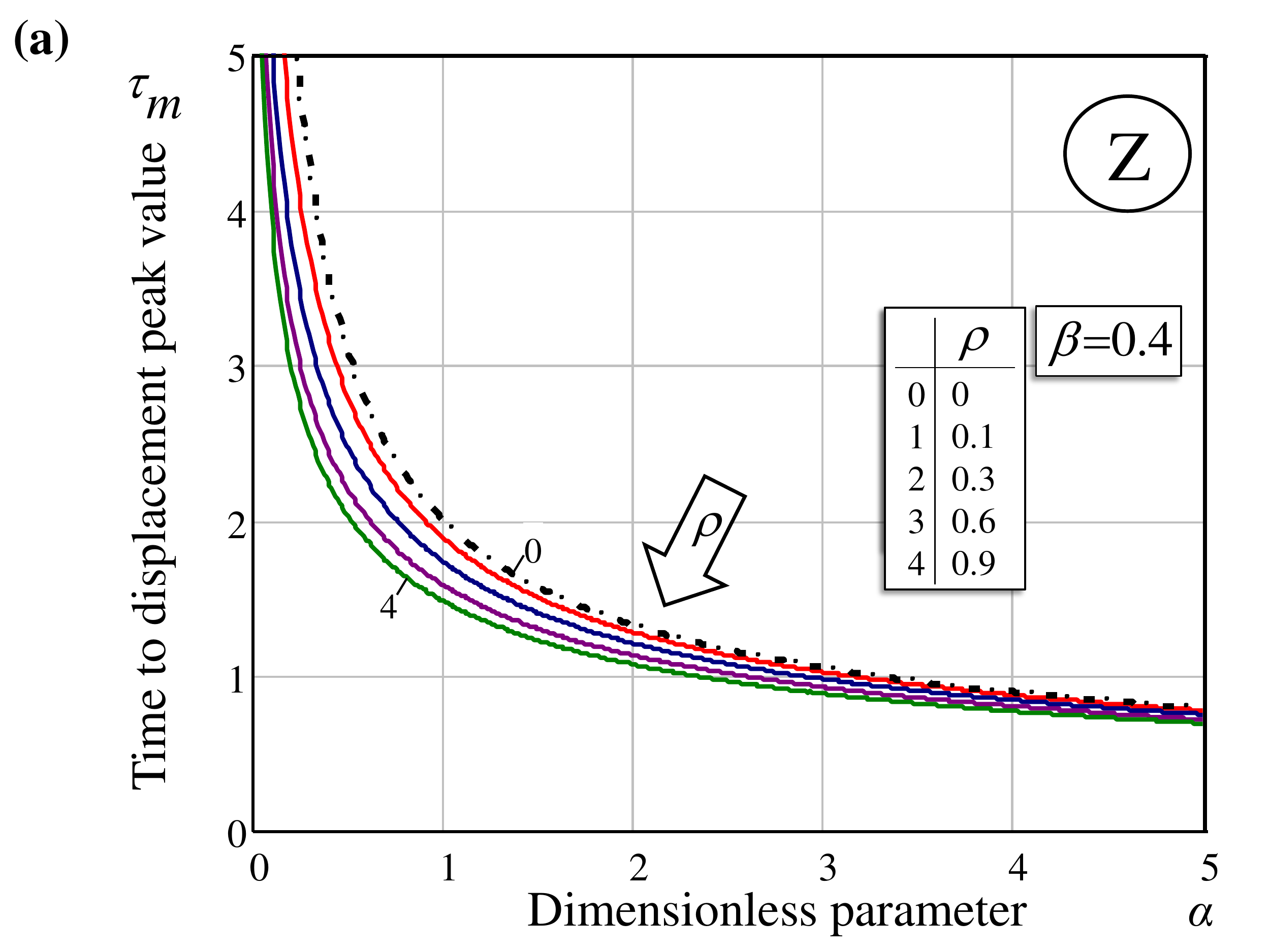}
    \includegraphics [scale=0.30]{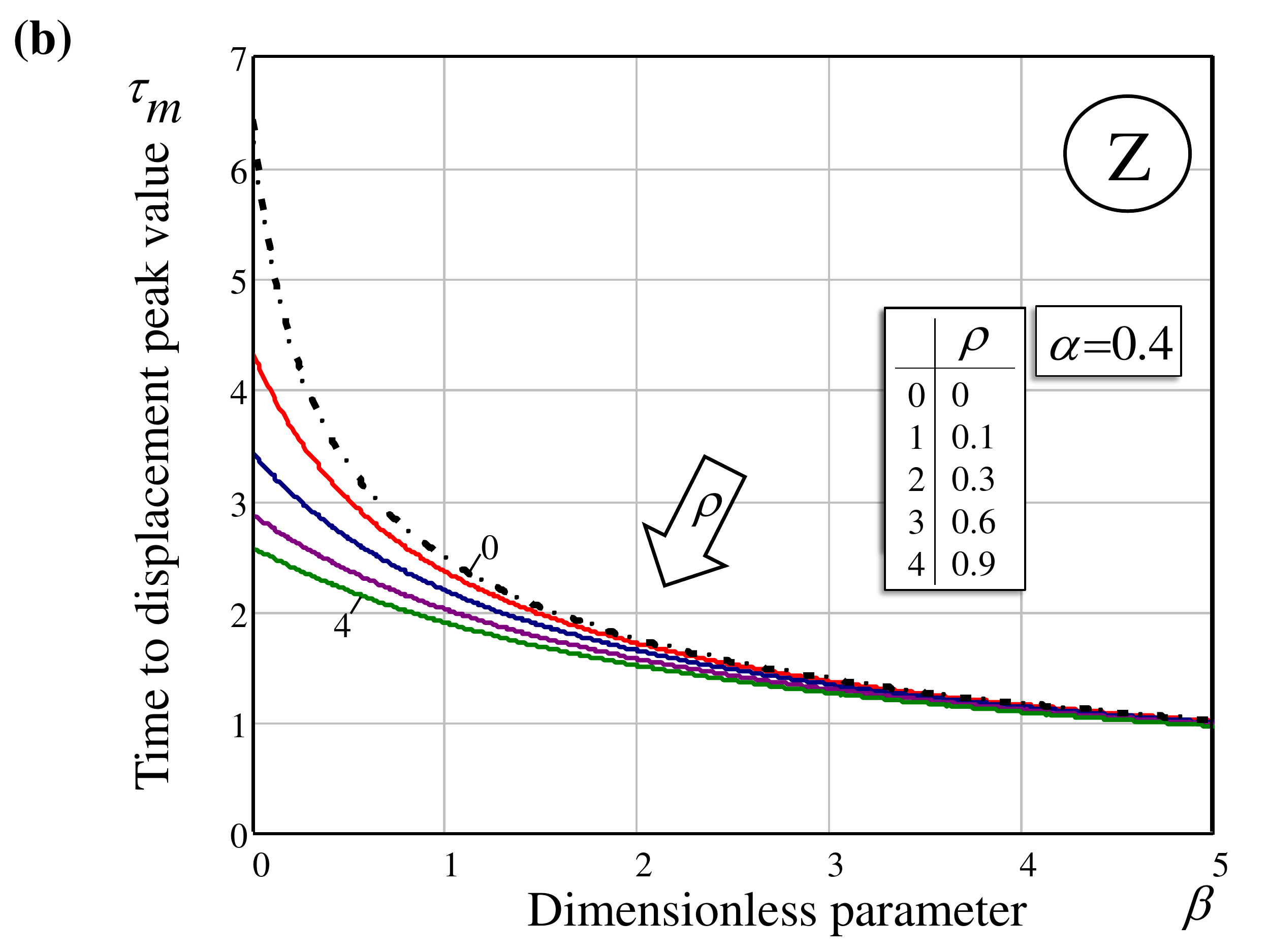}
        \caption{Time to displacement peak value $\tau_m=t_m/\tau_R$ in the QLV standard solid model as a function of:
(a) the reciprocal relative mass of the impactor $\alpha$ with the fixed relative incident velocity $\beta=0{.}4$ and
(b) the relative incident velocity $\beta$ with the fixed reciprocal relative mass of the impactor $\alpha=0{.}4$ for different values of the elastic moduli ratio $\rho$.}
    \label{taumQLVSSM=04}
\end{figure}

\begin{figure}[h!]
    \centering
    \includegraphics [scale=0.30]{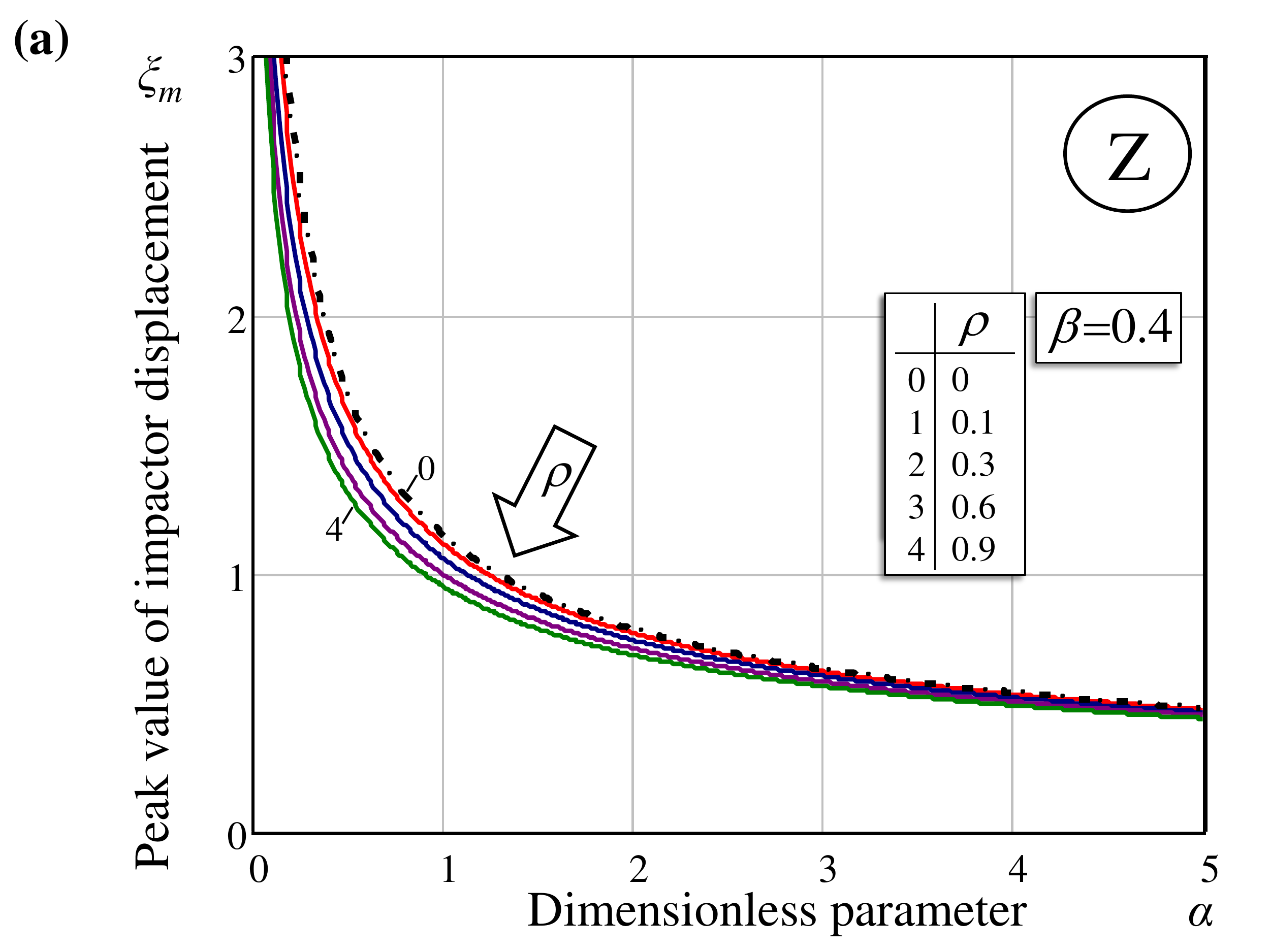}
    \includegraphics [scale=0.30]{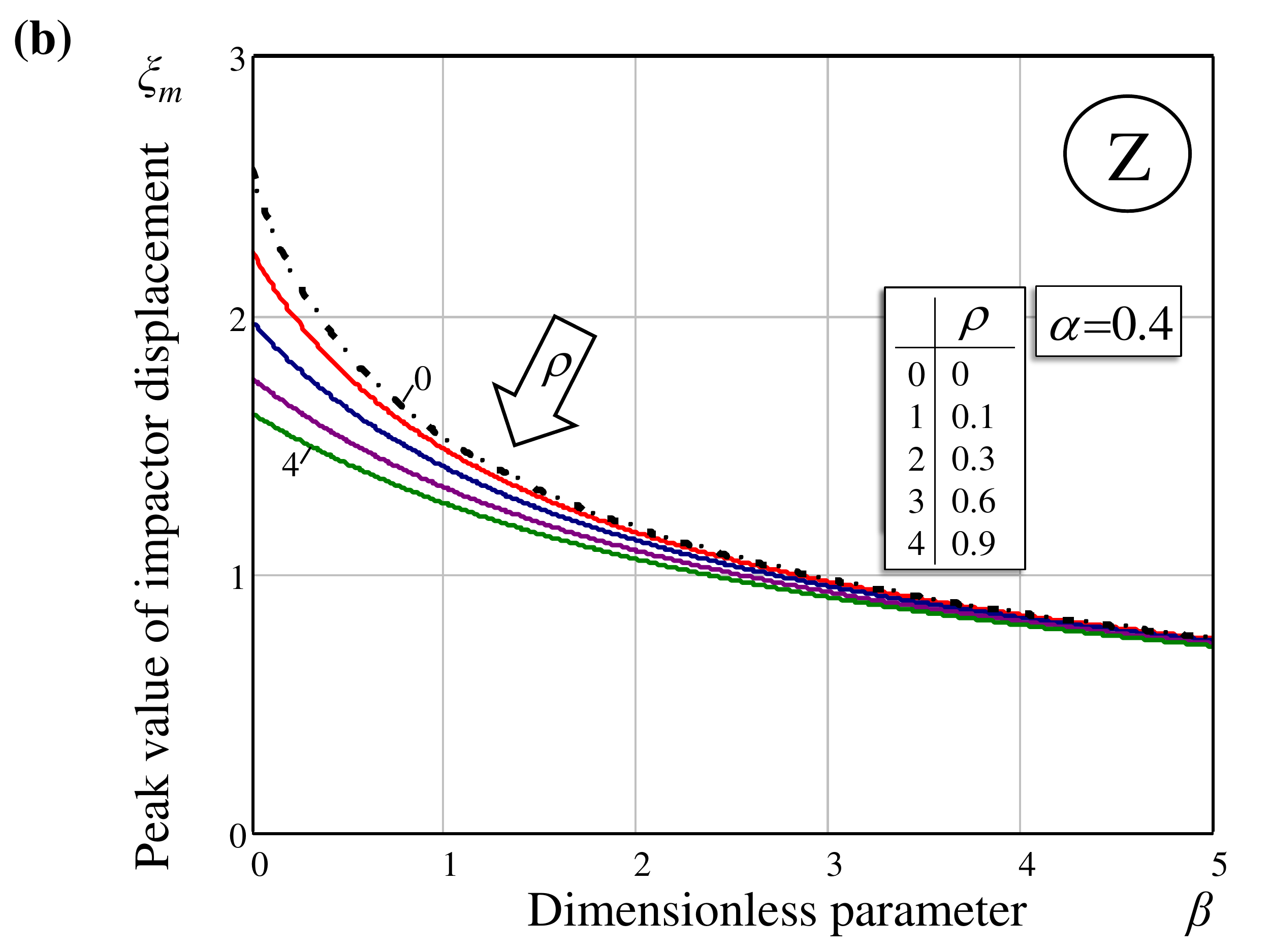}
        \caption{Peak value of impactor displacement $\xi_m(\tau_m)=x_m/(v_0\tau_R)$ in the QLV standard solid model as a function of:
(a) the reciprocal relative mass of the impactor $\alpha$ with the fixed relative incident velocity $\beta=0{.}4$ and
(b) the relative incident velocity $\beta$ with the fixed reciprocal relative mass of the impactor $\alpha=0{.}4$ for different values of the elastic moduli ratio $\rho$.}
    \label{ximQLVSSM=04}
\end{figure}

\begin{figure}[h!]
    \centering
    \includegraphics [scale=0.30]{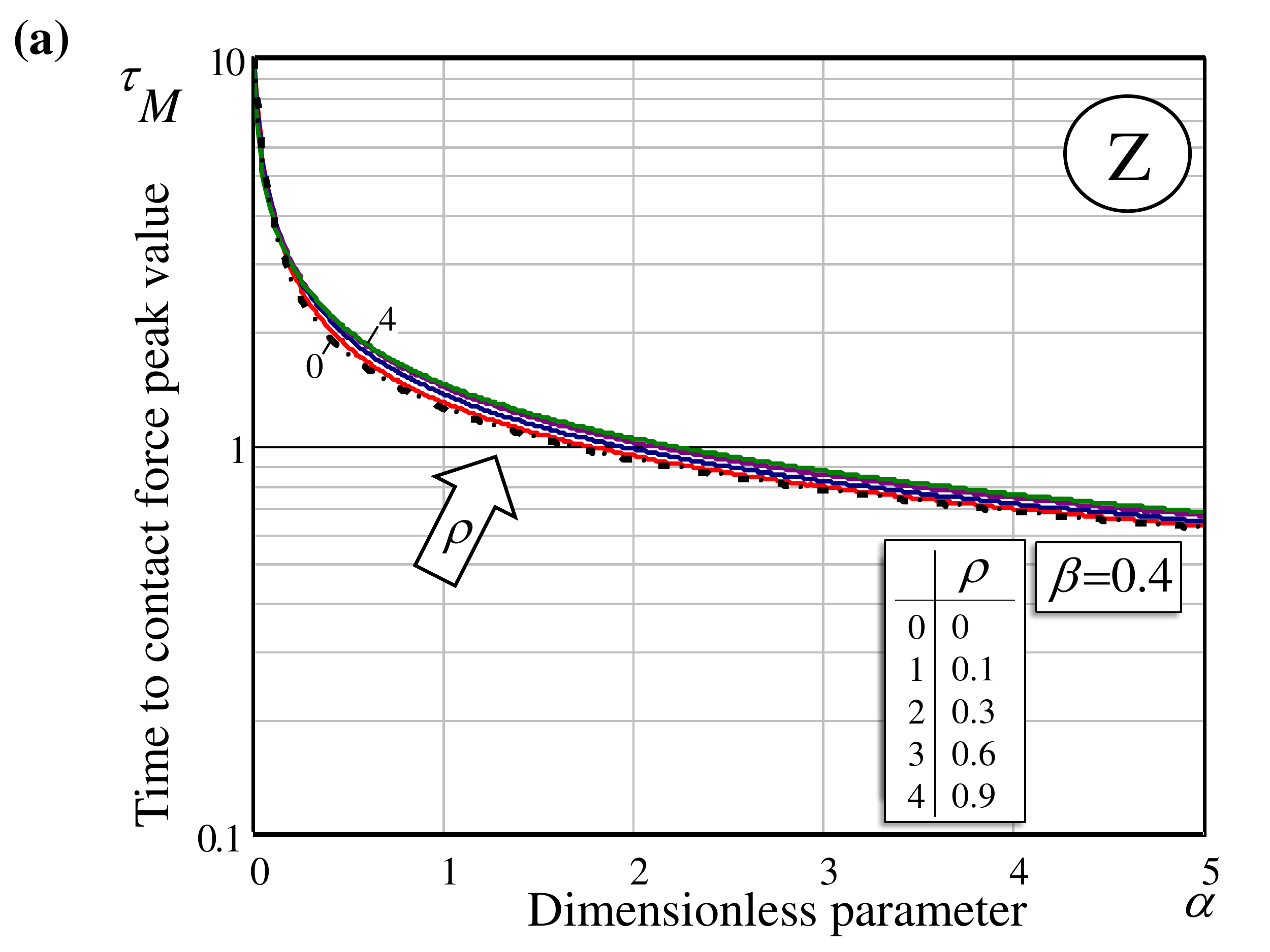}
    \includegraphics [scale=0.30]{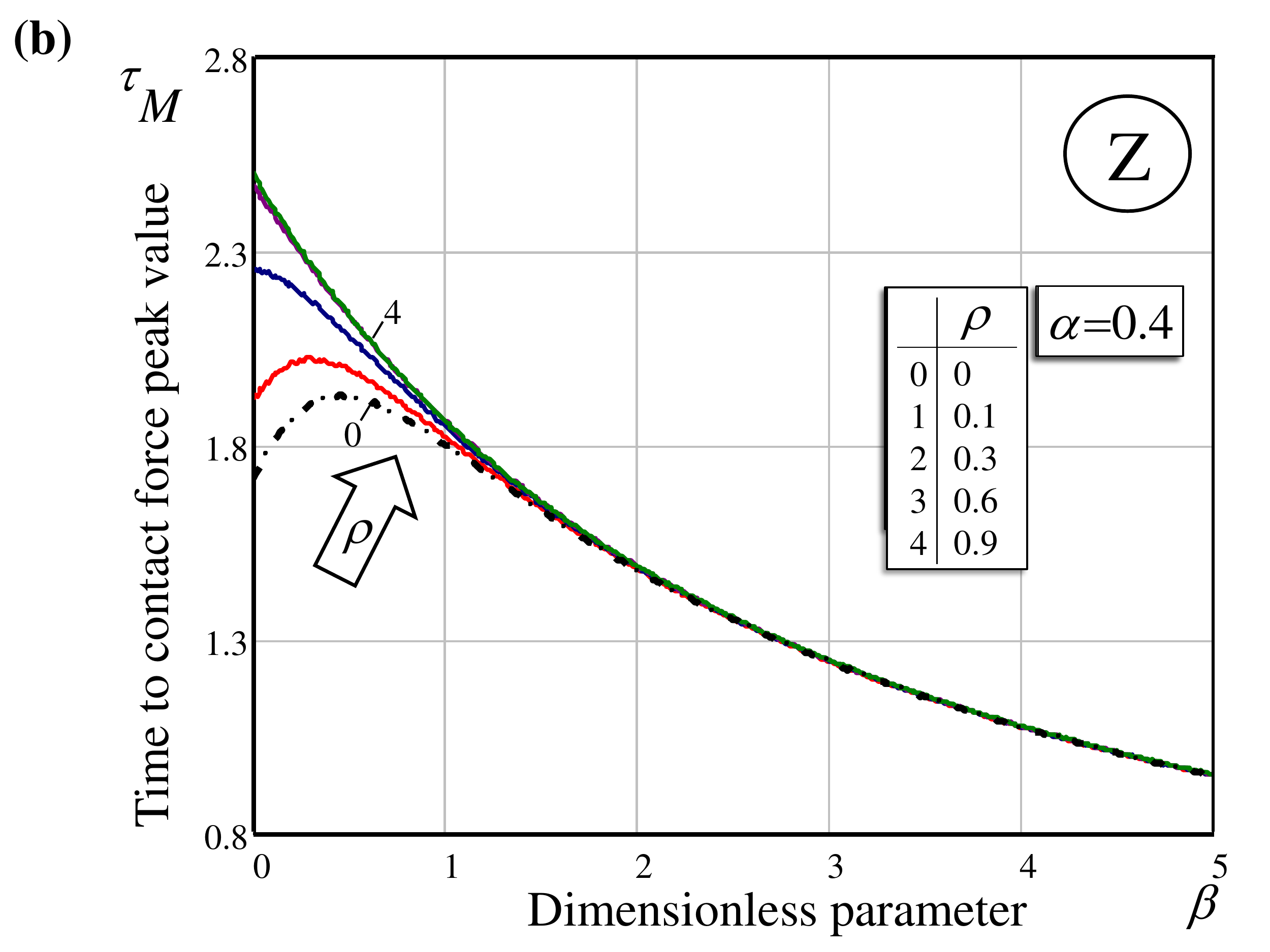}
        \caption{Time to contact force peak value $\tau_M=t_M/\tau_R$ in the QLV standard solid model as a function of:
(a) the reciprocal relative mass of the impactor $\alpha$ with the fixed relative incident velocity $\beta=0{.}4$ and
(b) the relative incident velocity $\beta$ with the fixed reciprocal relative mass of the impactor $\alpha=0{.}4$ for different values of the elastic moduli ratio $\rho$.}
    \label{tauMQLVSSM=04}
\end{figure}

\begin{figure}[h!]
    \centering
    \includegraphics [scale=0.30]{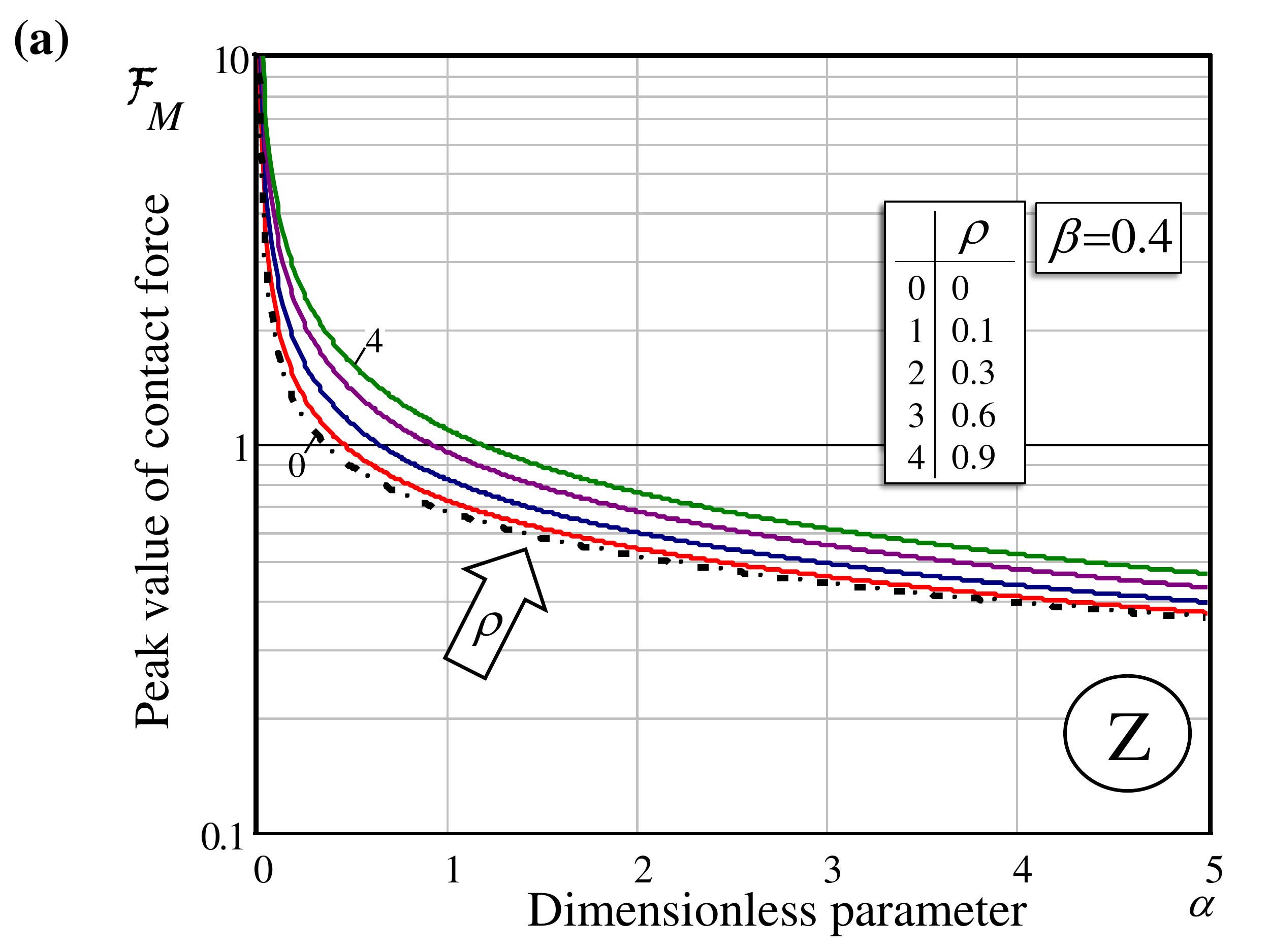}
    \includegraphics [scale=0.30]{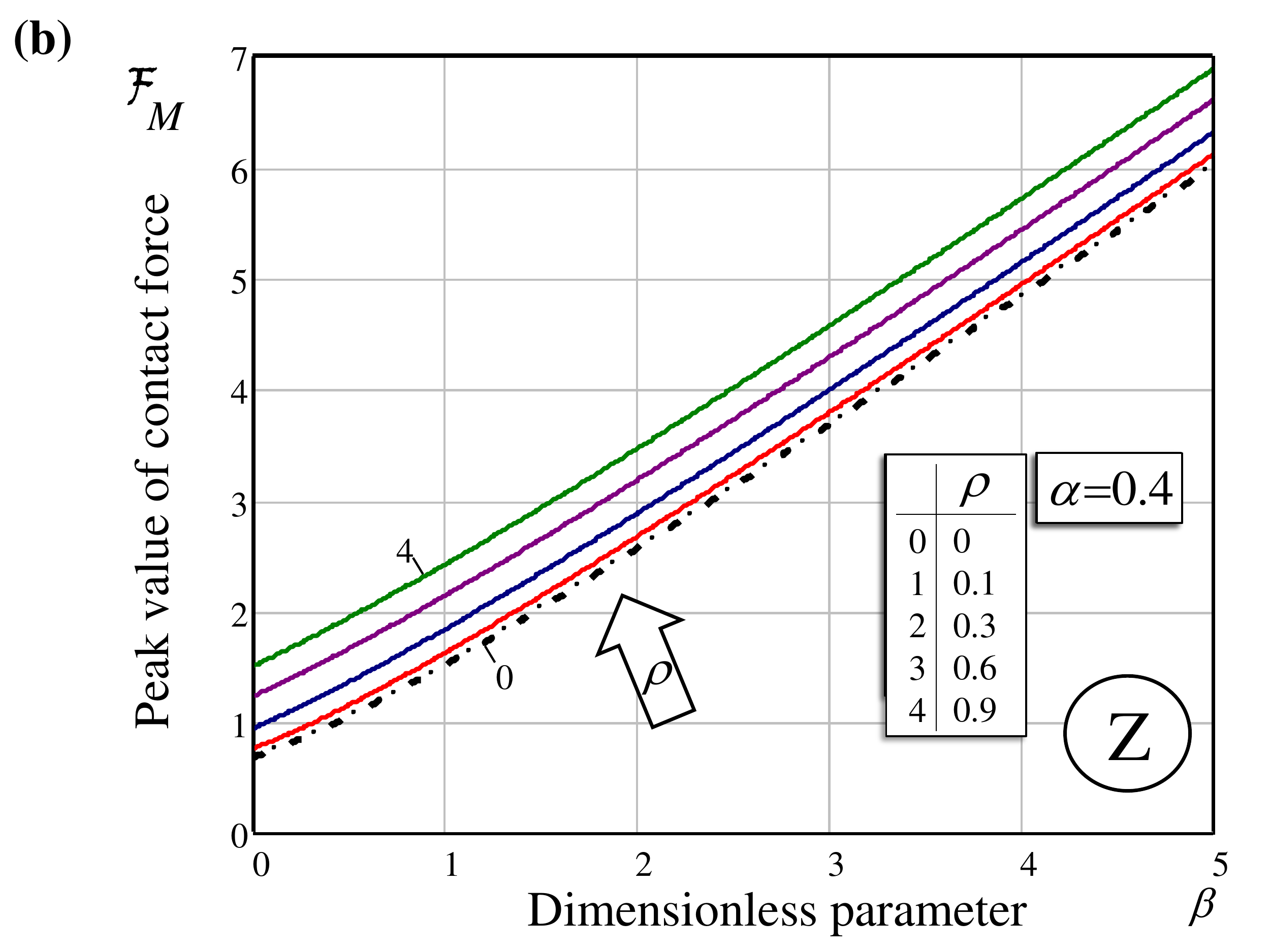}
        \caption{Peak value of contact force
$\mathcal{F}_{\rm M}\bigr\vert_{\tau=\tau_M}=F_M h_0/(AE_0)$ in the QLV standard solid model as a function of:
(a) the reciprocal relative mass of the impactor $\alpha$ with the fixed relative incident velocity $\beta=0{.}4$ and
(b) the relative incident velocity $\beta$ with the fixed reciprocal relative mass of the impactor $\alpha=0{.}4$ for different values of the elastic moduli ratio $\rho$.}
    \label{FQLVSSM=04}
\end{figure}

Further, the QLV Zener impact model contains an additional dimensionless parameter $\rho$.
The behavior of the main impact parameters for the QLV standard solid model is studied for the following different values of the elastic moduli ratio $\rho$:  $0{.}1$, $0{.}3$, $0{.}6$, $0{.}9$, and $0{.}0$, corresponding to the limiting case of QLV Maxwell model.
In Fig.~\ref{eQLVSSM=04}--\ref{FQLVSSM=04}, in the case (a) we use the range of reciprocal relative impactor mass $\alpha$ from 0 to 5 with the fixed relative incident velocity $\beta=0.4$, while in the case (b) the range of relative incident velocity $\beta$ from 0 to 5 is employed with the fixed reciprocal relative impactor mass $\alpha=0.4$.

As a result of numerical calculations, we observe a nonmonotonic behavior of the coefficient of restitution for the QLV Zener model. In Fig.~\ref{eQLVSSM=04} (a) and (b), the initial dropping part of each curve (except the limiting case of QLV Maxwell model) is followed by the ascending part of the curve after passing through the minimum point, which shifts to the right with increasing $\rho$.

The nonmonotonic of the coefficient of restitution is reflected by a non-trivial dependence of the relative contact duration on the parameter $\rho$ (see Fig.~\ref{taucQLVSSM=04}, where the curves cross over). Nevertheless, the relative contact duration decreases with the increase of either relative incident velocity or reciprocal relative impactor mass (see Fig.~\ref{taucQLVSSM=04}).

The behavior of the relative time to displacement peak value (see Fig.~\ref{taumQLVSSM=04}) and the relative peak value of impactor displacement (see Fig.~\ref{ximQLVSSM=04}) is similar and, in particular, the two quantities both decrease with increasing the elastic moduli ratio (the direction of the increasing
$\rho$ is shown by the wide arrow). Note that the dimensional peak value of impactor displacement is given by $x_m=v_0\tau_R\xi_m(\tau_m)$, therefore $x_m$ will increase with increasing $v_0$.

Fig.~\ref{tauMQLVSSM=04} shows that the time to contact force peak value increases with the increase of the elastic moduli ratio. At the same time, the peak value of contact force in the QLV standard solid model monotonically increases when either the impact mass $m$ or the incident velocity $v_0$ increases (see Fig.~\ref{FQLVSSM=04} (a) and Fig.~\ref{FQLVSSM=04} (b), respectively).

\section{Discussion and conclusion}
\label{1dsSectionDC}

First of all, let us compare the predictions of the two limit QLV models for same specific values of the dimensionless parameters. It is of interest that the two models exhibit different trends in regard to the coefficient of restitution. Namely, for the QLV Maxwell and Kelvin--Voigt models, the quantity $e_*$ increases and decreases, respectively, with increasing the impactor velocity (see Fig.~\ref{eQLVbeta.pdf}). At the same time, the dependences of the coefficient of restitution on the relative impactor mass are similar to those of the corresponding linear models (see Fig.~\ref{eQLValpha.pdf}).

It is clear that the nonlinear description of the elastic stress function proposed by \citet{Fung1981}
plays an important role in the qualitative behavior of the solution of the impact problem in the QLV Zener model. In particular, introducing the non-linearity parameter $B$, we make the behavior of the coefficient of restitution to be dependent on the impactor speed at incidence. Moreover, the choice of the stress-relaxation functions contributes to the model's qualitative performance as well.
Observe that, because both parameters $\beta$ and $\beta^\prime$ are proportional to $v_0$ and $B$, the effect of increasing non-linearity parameter $B$ on the coefficient of restitution $e_*$ will be the same as the effect of increasing $v_0$.

In the case under consideration, the coefficient restitution for the QLV Zener model in the range of small values of the incident velocity $v_0$ or the reciprocal relative impactor mass $1/m$ slightly decreases similar to the tendency observed in the case of QLV Kelvin-Voigt model, while for other considered values of the parameters $\beta$ and $\alpha$ the coefficient restitution for the QLV Zener model increases similar to the QLV Maxwell model.
It can be suggested that the crossing curves of contact duration observed in Fig.~\ref{taucQLVSSM=04} for the increasing elastic moduli ratio $\rho$ in the QLV standard solid model can shed a light on the behavioral transition from one pattern to the other, corresponding to the limiting cases. However, it should be taken into account that while the QLV Maxwell model is recovered from the QLV Zener model in the limit as $\rho\to 0$, the passage from the QLV Zener model to the QLV Kelvin-Voigt model requires a certain renormalization as it was shown in Section~\ref{1dsSub1section4}.

It was shown \citep{ArgatovMishuris2011AMM,Ardatov2013} that the short-time response of a linear biphasic layer, whose deformation is described in the framework of the asymptotic model given by \citet{AteshianLaiZhuMow1994}, in blunt impact is mathematically (approximately) equivalent to that of the Maxwell model. Hence, one can anticipate that applying the QLV Maxwell model to articular cartilage is more appropriate than the choice of of the QLV Kelvin--Voigt model. However, the validity of the QLV models for articular cartilage under physiological load magnitudes should be verified experimentally at least for some range of the model parameters.

Finally, we should point out that there are strong limitations of the assumption of quasilinear elasticity for soft biological tissues. In particular, it has been shown that when fitting the quasilinear viscoelastic law to relaxation curves, creep curves are reasonably well reproduced, while hysteresis under cyclic loading is grossly underestimated \citep{BischoffArrudaGrosh2004}. Note also
\citep{ProvenzanoLakesCorr2002} that one of the known major drawbacks of Fung's QLV model is that its time dependence is independent of strain, therefore predicting the same relaxation rate regardless of applied strain.

\section*{Acknowledgment}

The authors acknowledge support from the FP7 IRSES Marie Curie grant TAMER No 610547R.

\end{document}